\documentclass[11pt,onecolumn,a4paper,accepted=2026-04-09]{quantumarticle}

\newif\ifARXIV
\newif\ifUSEBBL

\ARXIVtrue

\USEBBLtrue

\pdfoutput=1
\usepackage[font={small}]{caption}
\usepackage{subfigure}
\usepackage{graphicx}
\usepackage[numbers,sort&compress]{natbib}
\usepackage{float}
\usepackage[normalem]{ulem}
\usepackage{mathrsfs} 
\usepackage{multicol}
\usepackage{cancel}
\usepackage{mathtools}
\usepackage{hyperref}
\usepackage{xcolor}
\usepackage{simplewick}
\usepackage{amsmath}
\usepackage{amsfonts}
\usepackage{amssymb}
\usepackage{graphicx}
\usepackage{simplewick}
\usepackage{hyperref}
\usepackage{multirow}
\usepackage{enumitem}
\usepackage{physics}
\usepackage{booktabs}
\usepackage{xcolor}
\usepackage{dsfont}
\usepackage{subfigure}
\usepackage{amsmath,amsthm,appendix,bm,graphicx,hyperref,mathrsfs,setspace,slashed,xcolor}

\def\Z2{ {\mathbb Z}_{2} }

\newcommand{\BRA}[1] {\langle  #1 |}
\newcommand{\CMATRIX}[1] {{\mathbb C}^{ #1 \! \times \! #1}}

\newcommand{\KET}[1] {| #1 \rangle}
\newcommand{\KETBRA}[1] {\left| #1 \right\rangle \! \left\langle  #1 \right|  }

\def\NUMBER{ N }
\def\VIVIANI{ \xi }
\def\SPINSQUEEZING{KitagawaUedaPRA1993,HaldSorensenPRL1999,KuzmichMandelPRL2000,WangSanders2003,JinKimPRL2007,MaWangPhysRep2011,DegenReinhardRMP2017,EsteveGrossNat2008,PezzeSmerziRMP2018,ChaiLaoPRL2020,XinChapmanPRXQ2022,XinBarriosPRL2023,230915353,240219429}
\def\ALLNQI{PhysRevLett.81.3992,150706334,BechmannPLA98,9802051,9803019,9811036,0309189,0502072,BrunPRL09,BennettPRL09,13033537,13030371,13107301,200907800,220613362,XuPRR22,211105977,240416288,240310102}
\def\NQIEXPT{ShullAtwoodPRL1980,GahlerKleinPRA1981,WeinbergPRL89,BollingerPRL89,PhysRevLett.64.2261,WalsworthPRL90,MajumderPRL90,190901608,ArnquistAvignonePRL2022,PolkovnikovGramolinPRL2023,BrozYouPRL2023}

\begin{document}

\title[accepted=2026-04-09]{From spin squeezing to fast state discrimination}

\author{Michael R. Geller}
\email{mgeller@uga.edu}
\affiliation{Department of Physics and Astronomy, University of Georgia, Athens, Georgia 30602, USA}
\affiliation{Center for Simulational Physics, University of Georgia, Athens, Georgia 30602, USA}
\date{April 30, 2026}

\begin{abstract}
There is great interest in generating and controlling entanglement in Bose-Einstein condensates and similar ensembles for use in quantum computation, simulation, and sensing. One class of entangled states useful for quantum-enhanced metrology are spin-squeezed states of $N$ two-level atoms.  After preparing a spin coherent state of width $ {N}^{-\frac{1}{2}} \! $ centered at coordinates $( \theta, \phi) $ on the Bloch sphere, atomic interactions generate 
an evolution that shears the state's probability density, stretching it to an ellipse and causing squeezing in a direction perpendicular to the major axis. Here we consider the same  setup but in the $N \rightarrow \infty $ limit  (simultaneously rescaling scattering lengths by $1/N$). This shrinks the initial coherent state to zero area. Large $N$ also suppresses two-particle entanglement and squeezing, as required by a monogamy bound. The torsion (1-axis twist) is still present, however, and the center $( \theta, \phi) $ of the large $N$ coherent state evolves as a qubit governed by a two-state Gross-Pitaevskii equation. The resulting nonlinearity is known to be a powerful resource in quantum computation. In particular, it can be used to implement single-input quantum state discrimination, an impossibility within linear one-particle quantum mechanics. In this problem, a qubit state $ | \psi \rangle \in \{ | a \rangle , | b \rangle \} $ is input to a procedure (with knowledge of the candidates $ | a \rangle , | b \rangle $) that  attempts to determine which was provided. We obtain a solution to this discrimination problem in terms of  a Viviani curve on the Bloch sphere. We also consider an open-system variant containing both Bloch sphere torsion and dissipation. In this case it should be possible to generate two basins of attraction within the Bloch ball, having a shared boundary that can be used for a type of autonomous state discrimination that is independent of  $ | a \rangle , | b \rangle $. We explore these and other connections between spin squeezing in the large $N$ limit and nonlinear quantum mechanics, and argue that a two-component condensate is a promising experimental platform for realizing a nonlinear qubit.
\end{abstract}

\maketitle
\setcounter{tocdepth}{2}
\clearpage
\tableofcontents


\section{Introduction}
\label{introduction section}

\subsection{Squeezing and nonlinear gates}
 
 A major theme in quantum sensing and 
metrology is the use of entanglement to improve the precision of parameter estimation beyond the standard quantum limit \cite{CavesThorneRMP1980,GiovannettiLloydSci2004,GiovannettiLloydPRL2006,\SPINSQUEEZING}. It has long been recognized that
quantum spin ensembles and
Bose-Einstein condensates (BECs) provide natural platforms for preparing and controlling the squeezed states sufficient for achieving this sensitivity enhancement \cite{\SPINSQUEEZING}. In the case of a two-component 
condensate with tunable atomic interaction, the ${\rm SU}(2)$ or spin coherent state \cite{RadcliffeJPA1971,CiracPRA98,ByrnesWenPRA12,14103602}
\begin{eqnarray}
\KET{ F}  := 
\frac{ ( \psi_0  a_0^\dagger + \psi_1 a_1^\dagger )^\NUMBER }{\sqrt{\NUMBER ! }} \,  \KET{{\rm vac}} , \ \ 
\psi_{0,1}  \in {\mathbb C} , \ \ 
\psi_0 = {\textstyle{ \cos( \frac{\theta}{2}) }}, \ \ \psi_1 = e^{i \phi} \, { \textstyle{ \sin( \frac{\theta}{2} )}} ,
\label{coherent state}
\end{eqnarray}
is prepared, centered at coordinates $( \theta, \phi) $ on the Bloch sphere.  The operator
 $ \psi_0  a_0^\dagger + \psi_1 a_1^\dagger $ creates an atom in a superposition of spin or
pseudospin components 0 and 1, and 
$\NUMBER$ is the total number of two-level atoms, which we call bosonic qubits. The 
$a_{\alpha} \, ( \alpha \in \{0,1\})$ 
are bosonic annihilation operators for the two internal states. 
The many-body state (\ref{coherent state}), parameterized by coordinates $( \theta, \phi) $, encodes a single qubit, because the same probability amplitudes are stored in each of the $\NUMBER$ identical bosons. The state is separable and is simplest to prepare with the interaction turned off. Atomic collisions then generate an interaction, such as the Kitagawa-Ueda one-axis-twist \cite{KitagawaUedaPRA1993}, which shears the state's probability density, stretching it to an ellipse and squeezing it in a direction perpendicular to the major axis \cite{\SPINSQUEEZING}.

\begin{figure}
\begin{center}
\includegraphics[width=7.0cm]{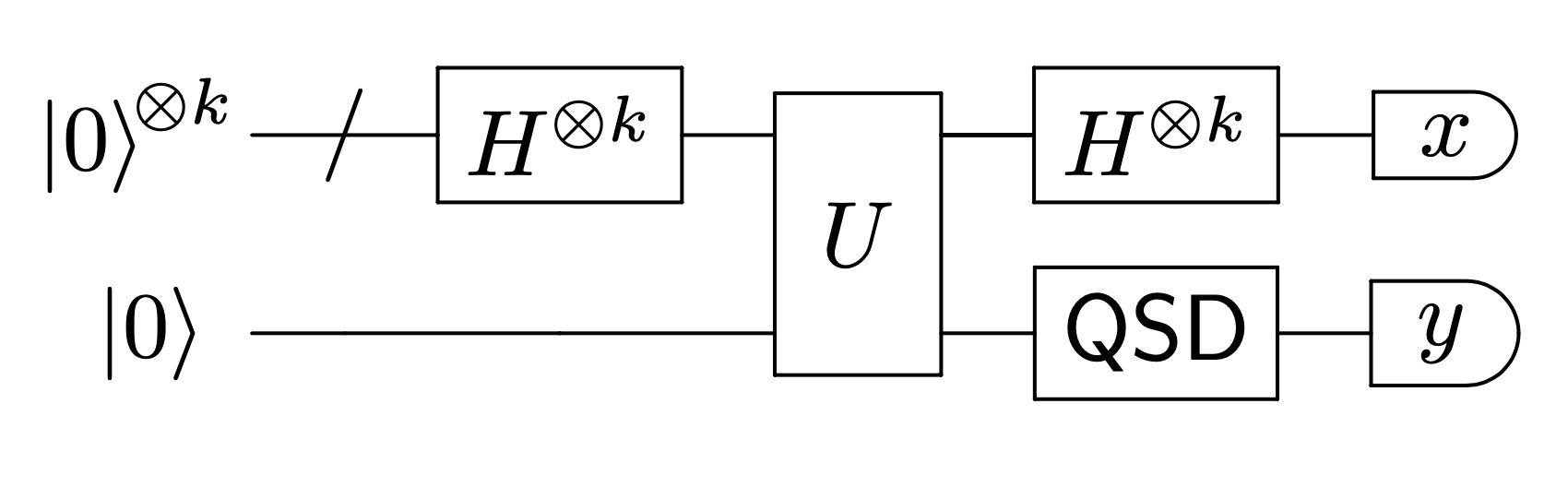} 
\caption{Efficient reduction of 3SAT to quantum state discrimination (QSD). Here $U$ implements the oracle and the first register is postselected to $\KET{0}^{\! \otimes k}$. The QSD gate implements fast discrimination by simulating a nonlinear qubit. The output of QSD is a classical state, $ \KET{0}$ or $\KET{1}$.} 
\label{npcomplete circuit figure}
\end{center}
\end{figure} 

In this paper we argue that the Bloch sphere torsion responsible for enhanced metrology through squeezing can also be used to enhance quantum computation. It is known that nonlinear quantum evolution in idealized models can significantly enhance information processing \cite{\ALLNQI}. This is true even if there is only {\it one} nonlinear qubit, acting as a coprocessor, coupled to a linear and otherwise conventional quantum computer \cite{PhysRevLett.81.3992,150706334,240310102}.  In this simplified setting, there is already an
efficient reduction from any problem in the complexity class $ \mathsf{NP} $---for which it is possible to efficiently check a proposed solution---to the problem of discriminating 
exponentially close single-qubit states \cite{PhysRevLett.81.3992,150706334}. By {\it efficient reduction} we mean a polynomial-depth quantum circuit  for solving one problem in terms of another. Discriminating between two possible states,
 $ \KET{a}$ and  $\KET{b} $, means  inputting $ \KET{\psi} \in \{ \KET{a} , \KET{b} \} $ into a procedure that determines whether $ \KET{\psi}$ is  $\KET{a}$ or $ \KET{b}$;  normally this requires many perfect copies of the input \cite{08101970,12042313,BaeJPA15,210812299}.

\subsection{Reduction of 3SAT to state discrimination}

To understand the reduction, note that it is sufficient to reduce the Boolean satisfiability problem 3SAT to state discrimination, because 3SAT is $\mathsf{NP}$-complete \cite{AroraBarak2009}. 
Each instance is 
specified by a Boolean function $f : \Z2^k \rightarrow\Z2 $ on $k$ bits, in conjunctive normal form (logical AND of clauses, with each clause the  OR of up to three variables or their negations).
 The task is to determine whether there is 
 at least one assignment of the $k$ 
variables for which the formula evaluates to \textsc{True}. If so, $f$ is satisfiable. 
The reduction given by Abrams and Lloyd \cite{PhysRevLett.81.3992} assumes an efficient gate decomposition for the {\it oracle} operator $U \KET{x} \KET{y} = \KET{x} \KET{y \oplus \! f(x)}$, as well as access to a fault-tolerant quantum computer. Here $\oplus$ denotes addition mod 2. 
  
The reduction circuit, shown in Figure~\ref{npcomplete circuit figure}, measures a global property of $f$ (number of satisfying assignments) by one application of the  oracle to a uniform superposition of classical states $\{ \KET{x} \}_{x \in \Z2^k} $. Initialize $k$ qubits and an ancilla qubit to $ \KET{0}^{ \! \otimes k} \KET{0} $.  Next apply Hadamard gates $H^{\otimes k}$ to the first $k$ qubits,  where $ H^{\otimes k} \KET{x} = 2^{-\frac{k}{2}} 
\sum_{x^\prime \in \Z2^k} (-1)^{ {\vec x} \cdot  {\vec x}^\prime } \, \KET{x^\prime} $, with
$ {\vec x} \cdot  {\vec x}^\prime =  \sum_i x_i x^\prime_i $.
This leads to
$ 2^{-\frac{k}{2}} \sum_{x^\prime  \in \Z2^k} \KET{x^\prime} \KET{0} $. 
Applying the oracle $U$ gives
\begin{eqnarray}
\frac{1}{\sqrt{2^k}}
\sum_{x \in \Z2^k} \KET{x} \KET{f(x)}.
\label{entangling step}
 \end{eqnarray}
A second application of  $H^{\otimes k}$
 then leads to
 \begin{eqnarray}
&&  \frac{1}{2^k}  \sum_{x \in \Z2^k}
 \sum_{x^\prime \in \Z2^k}   (-1)^{ {\vec x} \cdot  {\vec x}^\prime }
\KET{x^\prime}  \KET{f(x)} \\
 && = \KET{0}^{ \! \otimes k} \  
 \frac{ (2^k - s) \KET{0} +  s \KET{1}  }{2^k} 
 +  \frac{1}{2^k}  
  \sum_{x \in \Z2^k} \sum_{x^\prime \neq 0^k} (-1)^{  {\vec x} \cdot  {\vec x}^\prime  } 
\KET{x^\prime}  \KET{f(x)} ,
 \end{eqnarray}
where $s$ is the number of satisfying assignments. Measuring the first $k$ qubits
results in a classical output we call $x \in {\mathbb Z}_{2}^k $,  as indicated by the final measurement gate  in Figure~\ref{npcomplete circuit figure}. 
Postselecting for $ x=0$ gives
$ \KET{0}^{ \! \otimes k}  \KET{ {\rm out}(s) } $ with probability 
$  \frac{(2^k - s)^2 + s^2}{2^{2k}} 
\ge \frac{1}{2}$, where
\begin{eqnarray}
\KET{ {\rm out}(s) } :=  \frac{ (2^k - s) \KET{0} +  s \KET{1} }{ \sqrt{ (2^k - s)^2 + s^2 }} , \ \ 0 \le s \le 2^k.
 \end{eqnarray}
Distinguishing the cases $s=0$ ($f$ is not satisfiable) and  $s>0$ ($f$ is satisfiable)
is therefore equivalent to discriminating 
between potential ancilla states \cite{PhysRevLett.81.3992,150706334}
\begin{eqnarray}
\KET{a} = \KET{ 0} 
 \ \ {\rm and} \ \ 
 \KET{b} \in \big\{  \KET{ {\rm out}(1)}  , 
  \KET{ {\rm out}(2)}, \cdots ,  \KET{ {\rm out}(2^k)}\big\} .
 \label{discrimination problem for 3sat}
\end{eqnarray}
In this type of quantum state discrimination (QSD) gate, we are required to
determine whether the input state is $ \KET{ 0 } $, in which case the output returns $\KET{ 0 } $, or is equal to $ \KET{ {\rm out}(s) } $ for some $s > 0$,  
in which case the gate returns $\KET{ 1 } $.
But this reduction is not useful for linear quantum computers, however,  because in the hard case where there are only a few $ s \ll 2^k $ satisfying assignments, the states  $\KET{a}$ and $\KET{b}$ are very similar.  Their overlap
$ \langle  a |  b \rangle 
 = 1 -   \frac{s^2}{2} 2^{-2k} + O(2^{-3k}), \ 
 |\langle a | b \rangle|^2 =  1 -  s^2 2^{-2k} + O(2^{-3k}),$
when $k \gg 1$ and $ s \ll 2^k $, is exponentially close to unity. It is well known that discrimination protocols based on linear gates require exponential resources in this case \cite{08101970,12042313,BaeJPA15,210812299}, reflecting the limited information gained by measurement. Therefore, the reduction based on the circuit in Figure~\ref{npcomplete circuit figure} does not allow one to solve $\mathsf{NP}$-complete problems efficiently with a linear quantum computer. 

Abrams and Lloyd \cite{PhysRevLett.81.3992}  explored the implications of hypothetical forms of 
Schr\"odinger equation nonlinearity (which have not been observed experimentally \cite{\NQIEXPT}) on the power of quantum computers, and argued that even simple types would enable {\it single-input} QSD of exponentially close states, enabling polynomial-time solution of $\mathsf{NP}$-complete problems (and beyond). It should be emphasized that linear quantum computers are not expected to solve $\mathsf{NP}$-complete problems efficiently 
(factoring is in $\mathsf{NP}$ but is not $\mathsf{NP}$-complete) \cite{BennettBernstein9701001}. So it's interesting to investigate the power of nonlinearity further \cite{\ALLNQI}. 
An interesting series of papers assumed the Gross-Pitaevskii nonlinearity \cite{GrossNuovoCimento1961,PitaevskiiJETP1961} used to model ground states of 
dilute weakly interacting bosons in a mean-field approximation, finding speedups for both searching \cite{150706334,13033537,13030371,13107301,200907800} and QSD \cite{150706334}.
But specific experimental proposals for their implementation have not been available until recently \cite{240416288,240310102}, and none of the protocols has been demonstrated. 

The objective of this paper is to argue that the nonlinearity responsible for spin squeezing can also be harnessed for fast QSD. This is the 
subject of Sections~\ref{bloch sphere torsion and squeezing} and \ref{single-input state discrimination section}. In Section~\ref{torsion with dissipation section} we discuss a variation obtained by combining Bloch sphere torsion with dissipation, leading to a pair of basins of attraction within the Bloch ball that can be used to implement a form of  {\it autonomous} QSD. 
Section \ref{experimental realization section} discusses a possible realization of the torsion model with ultracold $^{39}$K atoms.
The nonlinear approach is summarized and its limitations are discussed in Section~\ref{conclusions section}.

\clearpage

We conclude the introduction with some comments 
on complexity (Section  \ref{trading time complexity for space complexity}), coupling to and entangling with linear qubits
(Section \ref{coupling linear and nonlinear qubits}),
and the large $\NUMBER$ limit and entanglement monogamy (Section  \ref{large n limit and 2-particle entanglement}). The main results are also summarized (Section \ref{main results}).

\subsection{Trading time complexity for space complexity}
\label{trading time complexity for space complexity}
     
The first comment regards speedup: Every BEC is an $\NUMBER$-body problem governed by linear quantum mechanics, so how can it provide nonlinear speedup? In the nonlinear approach, that same BEC is (approximately) described by a one-body problem with self-interaction, and the accuracy of the nonlinear picture requires $\NUMBER$ to be large (and the interaction weak). The atomic gas is used as a register of bosonic qubits initialized into a symmetric product state (\ref{coherent state}) via condensation, and subsequently controlled by applying fields and
varying the interaction. Time speedup is possible here because we have traded time complexity for space complexity: In the nonlinear picture there is only one logical qubit. The exponential time cost for discriminating exponentially close states with a linear quantum computer is transformed into a large $\NUMBER$ requirement. 

How large? The answer depends on the desired accuracy of the mean field theory. In a large family of condensate models (with $1/\NUMBER$ scaling of the interaction) it is possible to rigorously upper bound the {\it model error}  
$ \epsilon := \| \rho_{\rm eff}(t) - \rho_1(t) \|_1 $ by 
\begin{eqnarray}
 \epsilon \le 
c  \frac{e^{t/ t_{\rm ent} }-1 }{\NUMBER} , 
\label{model error bound}
 \end{eqnarray}
where $t$ is the evolution time (assuming the condensate is initialized to a symmetric product state at $t \! = \! 0$), and where $c$ and $t_{\rm ent}$ are positive constants (model-dependent parameters independent of $t$ and $\NUMBER$) \cite{ErdosJSP09,211209005}. Here $\rho_{\rm eff}$  is the mean field state, $\rho_1$ is the exact state traced over all qubits but one, and $ \| \cdot \|_1 $ is the trace norm  (Schatten 1-norm)
\begin{eqnarray}
\| X \|_1 :=  {\rm tr}( |X| ), \ \ |X| = \sqrt{X^\dagger X} .
\label{trace norm definition}
 \end{eqnarray}
The exponential growth of the worst case error  in (\ref{model error bound}) is to be expected, with a rate determined by the type, strength, and range of the interaction \cite{LiebCMP72}. The $1/\NUMBER$ dependence reflects two-particle entanglement monogamy (discussed below). We see from (\ref{model error bound}) that there is a short-time window $t < t_{\rm ent}$  where the required number of condensed atoms $ \NUMBER \approx  c / \epsilon $ is constant (for fixed $\epsilon$). This is because the BEC is initialized into a product state, and it takes a time $ t_{\rm ent}$ for the atomic collisions to produce entanglement.  
However for long computations, exponentially many atoms $ \NUMBER \approx (c/\epsilon)  e^{t / t_{\rm ent}} $ are required. Long computations would also require error correction, which is not addressed here.

\subsection{Coupling linear and nonlinear qubits}
\label{coupling linear and nonlinear qubits}

The second comment concerns entanglement between a scalable quantum computer and the nonlinear coprocessor. To solve a given 3SAT instance, the linear qubits are entangled with the logical qubit (encoded in a BEC) in step (\ref{entangling step}). This requires coupling a conventional qubit  to a condensate in a controllable fashion. 

One promising approach builds on advances combining cold atom magneto-optical traps or optical lattices with electromagnetic traps for ions 
\cite{IdziaszekCalarcoPRA2007,TomzaJachymskiRMP2019,JyothiEgodapitiyaRSI2019,KarpaAtoms2021,LousGerritsmaAdvances2022,ZipkesPalzerNat2010,SchmidtWeckesserPRL2020,GerritsmaNegrettiPRL2012,JogerNegrettiPRA2014,EbghaSaeidianPRA2019}.
Protocols have been proposed for
coupling BECs to trapped-ion qubits
and verifying  entanglement \cite{GerritsmaNegrettiPRL2012,JogerNegrettiPRA2014,EbghaSaeidianPRA2019}.
A recent proposal by Gro{\ss}ardt \cite{240310102} considers a neutral atom qubit coupled to a BEC.
Ions immersed in a BEC  have demonstrated sympathetic cooling of the ions \cite{ZipkesPalzerNat2010,SchmidtWeckesserPRL2020}, but condensate-ion entanglement has not been demonstrated experimentally. 

\subsection{Large $N$ limit and 2-particle entanglement}

\label{large n limit and 2-particle entanglement}

The final comment concerns the 
suppression of entanglement in the large 
$\NUMBER$ limit. The standard model for a scalar
 BEC
\cite{WiemanPritchardRMP1999,DalfovoGiorginiRMP1999,LeggettRMP2001,MorschOberthalerRMP2006} 
on which the Gross-Pitaevskii equation is based is that of a dilute Bose gas at zero temperature 
with the short-range atomic interaction replaced by a contact interaction: 
\begin{eqnarray}
V_{\rm int}( {\bf r} -  {\bf r}^\prime) \approx U_{0} \, \delta({\bf r} -  {\bf r}^\prime) , \ \  U_{0}  = \frac{4 \pi \hbar^2 a_{\rm s}}{m} ,
\label{contact interaction}
\end{eqnarray}
where $a_{\rm s}$ is the s-wave scattering length and $m$ is the atomic mass. The use of the contact interaction requires low energy and dilute neutral atoms such that
$ {\bar r} \gg  | a_{\rm s} | $,  where ${\bar r}$ is the average inter-particle distance. In addition, the Gross-Pitaevskii equation is a mean field approximation, so $\NUMBER \gg 1$ is also required to suppress fluctuations. Thus, as is well known, the use of the Gross-Pitaevskii equation to model a BEC requires both $  {\bar r} \gg  | a_{\rm s} | $ and $\NUMBER \gg 1$, as well as low energy and temperature \cite{DalfovoGiorginiRMP1999}.
However the Gross-Pitaevskii description can become invalid in the $\NUMBER \rightarrow \infty$ limit (even while the gas remains dilute). This limit is especially problematic in multi-component BECs, where only the lowest energy (spatial or internal) modes are included in a model, the others neglected on account of their higher energy. But with $a_{\rm s}$ fixed, the ground state energy per particle can diverge as $\NUMBER\rightarrow \infty$, invalidating the low energy assumptions. To avoid this, we adopt the approach of rigorous investigations of the Gross-Pitaevskii equation by simultaneously rescaling scattering lengths by $1/\NUMBER$ as $\NUMBER \rightarrow \infty $
\cite{LiebSeiringerPRA2000,BardosMAA2000,200605486,FrohlichCMP07,ErdosPRL2007,RodnianskiCMP09,KnowlesCMP10,ChenLeeJSP11,14116284,PicklRMP15,Benedikter2016,FrohlichAM19,BrenneckeAPED19,191014521}, making 
the large $\NUMBER$ limit well defined. In particular, the error bound (\ref{model error bound}) assumes this rescaling of the interaction. 
Therefore, the nonlinear picture becomes exact under the four conditions: 
(i) $ T \rightarrow 0 $,
(ii) $ \NUMBER \rightarrow \infty $,
(iii) $ | a_{\rm max}  | \NUMBER^\frac{1}{3}  \rightarrow 0 $ (low density),
and 
(iv) $ a_{\rm max}  N  \rightarrow 
{\rm constant \,}$(weak coupling).
Here $a_{\rm max}$ is the largest (in magnitude) of the s-wave scattering lengths 
$ a_{00}, a_{01},$ and $a_{11}$
used below. 

The $1/\NUMBER$ dependence in (\ref{model error bound}) is anticipated from a 
monogamy inequality \cite{CoffmanPRA00,OsborneVerstraetePRL2006} for pairwise pseudospin entanglement when applied to a system with permutation symmetry. 
To see this, first note that  
$ \| \rho_{\rm eff}(t) - \rho_1(t) \|_1 
= |  {\vec r}_{\rm eff} - {\vec r}_1 | $
measures both unitary and 
nonunitary errors in the mean field state $\rho_{\rm eff} $, by which we mean differences in the Bloch vector orientations and lengths.
However due to the symmetry of the interaction  (\ref{contact interaction}) the error here is mainly nonunitary, caused by the pairwise (and higher order) entanglement of bosonic qubit 1 with $2, 3, \cdots \! ,  \NUMBER$ and its resulting entropy increase. 
(With the translational degrees of freedom of the atoms condensed into a single trap mode, the
only entanglement here is between the 
internal states of the atoms.)
So we can assume that the model error $\epsilon$ is predominantly a nonunitary error caused by  pseudospin entanglement. Let 
$ \tau( \rho_{ij} ) $ be the {\it tangle} (concurrence squared)  between bosonic  qubits $i$ and $j$, a measure of their shared entanglement detected via the 
two-qubit density matrix $ \rho_{ij} $, traced over other atoms. The tangle satisfies the monogamy inequality \cite{CoffmanPRA00,OsborneVerstraetePRL2006} 
\begin{eqnarray}
 \tau( \rho_{12} )  +  \tau( \rho_{13} ) + \cdots +  \tau( \rho_{1\NUMBER} ) \le 4 \, {\rm det}(\rho_1)  , \ \ \NUMBER \ge 3.
 \label{monogamy}
\end{eqnarray}
The left side is a sum of the tangles (each quantifying  two-qubit entanglement) between 
bosonic qubits 1 and 2, between 1 and 3, and so on.
The right side is a measure of the total entanglement of qubit 1 with the ensemble, including genuine multiparticle entanglement, detected via the reduced density matrix $\rho_{1}$ (note that every qubit state 
has  $ {\rm det}\rho \le \frac{1}{4}$). Because the system has permutation symmetry, the tangles are equal and satisfy an intuitive bound
\begin{eqnarray}
 \tau( \rho_{ij} )  \le \frac{1}{\NUMBER-1} .
 \label{intuitive tangle bound}
\end{eqnarray}
This has the same large $\NUMBER$ behavior as the model error bound (\ref{model error bound}). As previously noted by Yang \cite{YangPLA2006}, the pairwise entanglement (as measured by concurrence or tangle) in any permutation-symmetric state always vanishes in the large $\NUMBER$ limit:
\begin{eqnarray}
\lim_{\NUMBER \rightarrow \infty} \,  \tau( \rho_{ij} ) = 0.
 \label{monogamy limit}
\end{eqnarray}

In the nonlinear approach to quantum information processing, pseudospin entanglement is intentionally suppressed this way. Although the 
proof of inequality (\ref{monogamy}) for general $\NUMBER$ is somewhat complicated and took several years to complete 
\cite{OsborneVerstraetePRL2006,YangPLA2006} after the initial proof for 
$\NUMBER \! = \! 3$ \cite{CoffmanPRA00}, a simpler derivation is possible here due to the assumed permutation symmetry; this is provided in Appendix \ref{entanglement monogamy section}. 
For explicit calculations of concurrence in a variety of symmetric multiqubit states, including squeezed states, see Wang and M{\o}lmer \cite{WangMolmerEPJD2002} and 
Wang and Sanders \cite{WangSanders2003}.

\subsection{Summary of results}
\label{main results}

This paper is mainly a proposal for experiments at the interface of atomic physics, nonlinear quantum mechanics, and quantum information. The goal of the paper is to explain why they would be interesting, and to offer conventional (if somewhat idealized) microscopic models for their description.
The connections explored here between spin squeezing, expansive dynamics, and nonlinear gates is likely to be applicable beyond the specific physical platform and squeezing mechanism considered here. The main contributions are:
\begin{enumerate}

\item[(i)]

We propose to use an ultracold gas of $\NUMBER$  neutral two-level atoms to implement  a logical qubit evolving nonlinearly in time according to the Gross-Pitaevskii equation, with the accuracy controlled by $\NUMBER$. Atom-atom entanglement is intentionally suppressed by making $\NUMBER$ large, using bound (\ref{intuitive tangle bound}), and the interactions weak. Speedup of QSD is possible  because the exponential time cost is replaced by an exponential space cost. To enhance the power of a scalable quantum computer, the nonlinear coprocessor is entangled with a register of conventional qubits.

\item[(ii)]

We obtain conservation laws satisfied by the torsion model, and use them to design a 
nonlinear QSD gate based on Viviani’s curve (Section \ref{single-input state discrimination section}). After the gate, input $\KET{a}$ is mapped to  $\KET{0}$, and $\KET{b}$ is mapped to 
$\KET{1}$.

\item[(iii)]

We consider a dissipative torsion model containing a phase where the Bloch ball develops two basins of attraction  ${\mathbb V}_{+}$ and ${\mathbb V}_{-}$ flowing to opposing fixed points $ {\bf r}^{\rm fp}_{+} , {\bf r}^{\rm fp}_{-} $. We propose to use this dynamical system to implement an autonomous QSD gate that maps every input state in $ {\mathbb V}_{\! +}$ to  ${\bf r}^{\rm fp}_{+}$, and every state in $ {\mathbb V}_{\! -}$ to  ${\bf r}^{\rm fp}_{-}$ (Section \ref{torsion with dissipation section}).

\end{enumerate}

\section{Bloch sphere torsion and squeezing}
\label{bloch sphere torsion and squeezing}

\subsection{Trace distance monotinicity}

\begin{figure}
\begin{center}
\includegraphics[width=6.0cm]{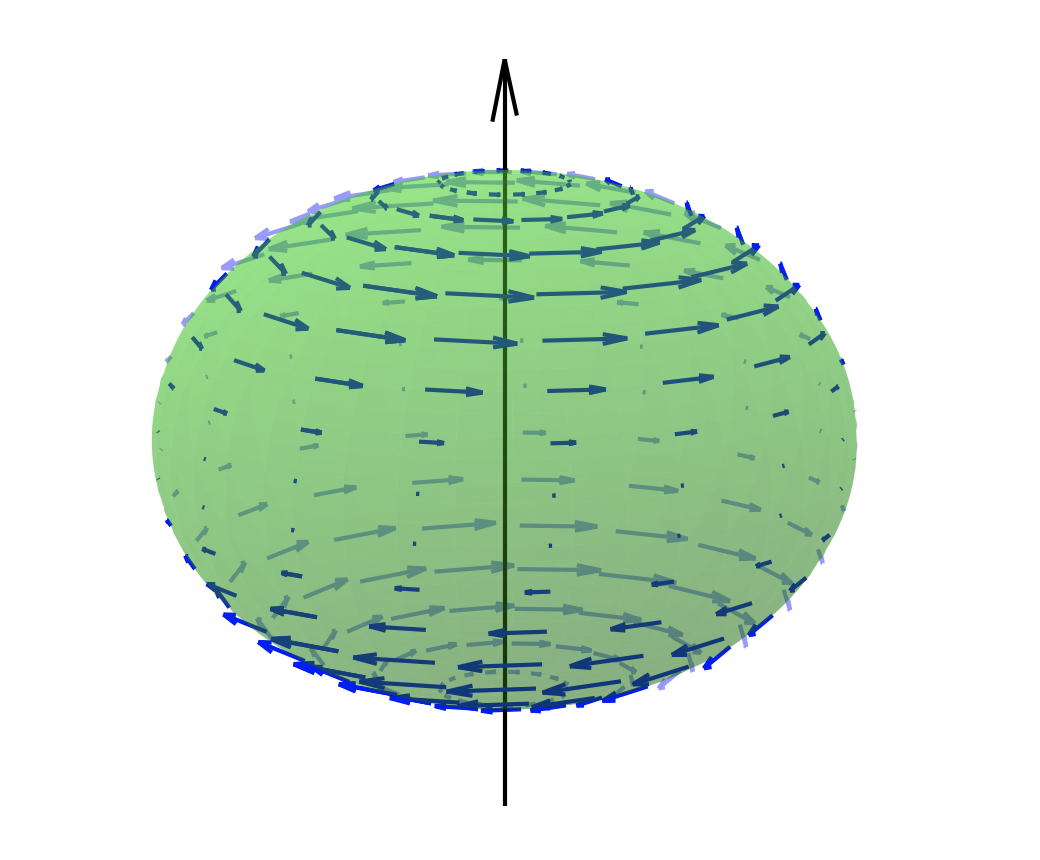} 
\caption{Block ball subjected to  $z$-axis torsion. All states in the upper hemisphere are expansive with all states in the lower hemisphere, a powerful quantum information processing resource.}
\label{z torsion figure}
\end{center}
\end{figure} 

To understand the importance of torsion and related concepts in nonlinear qubit models, it is useful to clarify precisely how nonlinear evolution
might benefit information processing. After all, single-qubit operations such as state preparation, unitary gates, and readout are routinely demonstrated with very high precision. Nonlinearity appears to offer no benefit for these operations \cite{MielnikJMP80}. Instead, consider how the entire state  space (the Bloch ball) flows 
under some continuous evolution, such as that generated by the Gorini-Kossakowski-Sudarshan-Lindblad  master equation \cite{GoriniJMP76,LindbladCMP76}. 
This global perspective focuses on the trajectories of a given state together with those nearby. 

Unitary gates rotate the Bloch ball rigidly around a fixed axis. We can say that unitary evolution on a pair of states $( \rho_{\rm a} , \rho_{\rm b} )$ is distance-preserving, where the distance $ \| \rho_{\rm a} - \rho_{\rm b} \|_1 $ is measured in trace norm
(for qubits with Bloch vectors $ {\vec r}_{{\rm a}}$ and $ {\vec r}_{{\rm b}}$ we note that 
$ \| \rho_{\rm a} - \rho_{\rm b} \|_1 =
{\rm tr} | \frac{ ( {\vec r}_{\rm a} - {\vec r}_{\rm b} )\cdot {\vec \sigma} }{2} |  =  |{\vec r}_{\rm a} - {\vec r}_{\rm b}| $). 
Dissipation and decoherence eventually map all states to a ground or mixed state. So nonunitary evolution is always contractive: quantum states (in the space of density matrices) move closer over time. But time evolution never moves a pair of states apart or decreases their overlap. 
This restriction is captured by the {\it monotonicity} of trace distance \cite{WildeQuantumInformation} (or relative entropy \cite{RuskaiRMP1994})
under linear time evolution: Let $ \rho_{\rm a}, \rho_{\rm b}  \in \mathbb{C}^{d \times d} $ be states (positive semidefinite  operators with unit trace), and let $\phi$ be a {\it linear} positive trace-preserving channel (complete positivity is not required). Then
\begin{eqnarray}
\| \phi( \rho_{\rm a}) - \phi(\rho_{\rm b})   \|_1 \le \| \rho_{\rm a} - \rho_{\rm b} \|_1 .
\label{trace distance monotinicity}
\end{eqnarray}
The trace norm is physically relevant 
because the normalized trace distance
 \begin{eqnarray}
D( \rho_{\rm a} , \rho_{\rm b}) := \frac{  \| \rho_{\rm a} -  \rho_{\rm b}  \|_1 }{2}
=  \max_{ 0 \preceq E \preceq I}  {\rm tr}[( \rho_{\rm a} \! - \!  \rho_{\rm b}) E ] , \ \  
0 \le D( \rho_{\rm a} , \rho_{\rm b}) \le 1,
 \end{eqnarray}
quantifies the distinguishability of  
$( \rho_{\rm a}, \rho_{\rm b})$ by any single (projective or POVM) measurement operator
$E \in \mathbb{C}^{d \times d} $ \cite{WildeQuantumInformation}.
The standard reference for (\ref{trace distance monotinicity}) is Ruskai \cite{RuskaiRMP1994}, who proved a more general monotonicity condition that includes (\ref{trace distance monotinicity}) as a special case (but assumes complete positivity).
Additional discussion of trace distance monotinicity and a direct proof of (\ref{trace distance monotinicity}) are  
provided in Appendix \ref{trace distance monotinicity section}.

\subsection{Nonlinear qubit with torsion}

A variety of nonlinear evolution equations, however, are known to violate trace distance monotonicity \cite{PhysRevLett.81.3992,150706334,211105977}, enabling a potential computational advantage that we explore here. A popular model for single-qubit nonlinearity is the $z$-axis torsion model \cite{150706334,211105977,240219429}
proposed by Mielnik \cite{MielnikJMP80} in 1980:
\begin{eqnarray}
\frac{ d \rho}{dt} = -i [ H_{\rm eff} , \rho] , \ \  H_{\rm eff} = g \, {\rm tr}(\rho \sigma^z) \sigma^z \! , \ \ \rho, H_{\rm eff}  \in {\mathbb C}^{2 \times  2} .
\label{torsion hamiltonian}
\end{eqnarray}
This Hamiltonian generates $z$ rotation with a frequency $2 g \, {\rm tr}(\rho \sigma^z)$ proportional to the $z$ coordinate of the state, as illustrated in Figure~\ref{z torsion figure}.
Under this evolution an initial Bloch vector
$ {\vec r} = {\rm tr}(\rho {\vec \sigma})$ is subjected to a state-dependent rotation, a feature absent in the SU(2) gate set. 
In particular,
states in the upper hemisphere in Figure~\ref{z torsion figure}
rotate in the opposite direction to those in the lower hemisphere. Pairs of such states can move apart from each other and violate trace distance monotinicity. Torsion about $x$ and $y$ are similarly generated by  $ {\rm tr}(\rho \sigma^x) \sigma^x $ and  $ {\rm tr}(\rho \sigma^y) \sigma^y $.   Luo {\it et al.} recently implemented these models, as well as two-axis
counter-twisting, with $^{87}$Rb atoms in an optical cavity \cite{240219429}. 

The expansivity in the torsion model
(\ref{torsion hamiltonian}) can be used to implement single-input QSD \cite{PhysRevLett.81.3992,150706334,211105977}, an impossibility in linear one-particle quantum mechanics \cite{08101970,12042313,BaeJPA15,210812299}. 
We will assume that $g \in {\mathbb R}$ is experimentally controllable and that linear Hermitian operators can be added to $H_{\rm eff}$ to implement arbitrary SU(2) gates.
The torsion model (\ref{torsion hamiltonian}) 
is not microscopic, but emergent, arising from a family of distinct microscopic models in their large $\NUMBER$ limits.
 In a recent paper \cite{240416288},
we mapped an atomtronic SQUID \cite{RyuSamsonNat2020,KapalePRL05,RyuPRL07,RamanathanWrightPRL2011,EckelNat14,KimZhuPRL2018,231105523}  
to the torsion model and used it to design a  single-input QSD gate for a logical qubit encoded in a rotating toroidal condensate. In this work we map a trapped two-component atomic BEC to the torsion model, and design fast discriminators for this setup as well.  A potential advantage of the two-component realization is that spin squeezing is now highly advanced and demonstrated by many labs worldwide \cite{\SPINSQUEEZING}. Another is pedagogical: In the two-component BEC 
(with no entanglement between translational and internal motion of the atoms), 
torsion dynamics can be obtained exactly from the Kitagawa-Ueda spin model \cite{KitagawaUedaPRA1993} in the large spin limit, confirming the mean field prediction  (Section \ref{kitagawa-ueda model large n}).

\subsection{Microscopic model}

In this section, we briefly review a two-component BEC model and use it to recover known spin squeezing results. Then in Section~\ref{path integral and large n limit} we take the large $\NUMBER$ limit and obtain the torsion model (\ref{torsion hamiltonian}). 
We consider a dilute condensate of trapped bosonic atoms of mass $m$ and metastable internal states $ \KET{\Psi_{0} } ,  \,
\KET{\Psi_1 }$ at  zero temperature  \cite{CiracPRA98}:
 \begin{eqnarray}
 H \!  = \!  \sum_{\alpha=0,1} \!  \int \! \! d^3r   
\bigg( \!  \phi^\dagger_\alpha {T}_{\alpha} \phi_\alpha
  \! +  \!   \frac{U_{\alpha \alpha}}{2} \, \phi_{\alpha}^\dagger  \phi_{\alpha}^\dagger \phi_{\alpha}  \phi_{\alpha} \!  \bigg)
 +   U_{01} \!  \! \int  \! \!  d^3r  \, \phi_{0}^\dagger  
 \phi_{1}^\dagger \phi_{0}   \phi_{1} 
 +    \frac{\Omega}{2}  \!  \! \int  \!  \! d^3r  \,
(\phi_{0}^\dagger  \phi_{1} \! + \!  \phi_{1}^\dagger  \phi_{0}), \ \
\label{two boson hamiltonian}
 \end{eqnarray}
 where $T_\alpha \! = \! - \frac{\nabla^2}{2m} + V_{\alpha}({\bf r}) $ is a single-atom Hamiltonian with state-dependent trapping potential  
$ V_{\alpha}({\bf r}) = \frac{1}{2} m \omega_{\alpha}^2 r^2 \! . $ Here
$ [\phi_{\alpha}({\bf r}) , \phi_{\alpha'}^\dagger({\bf r}') ] = \delta_{\alpha \alpha'}  \,\delta({\bf r}-{\bf r}') $, where $ \phi^\dagger_{\alpha}({\bf r}) $ acting on the vacuum creates an atom centered at position ${\bf r}$ in state $\KET{\Psi_\alpha }$. 
We also assume that $\langle \Psi_0 | \Psi_1 \rangle \! = \! 0$. Interaction parameters $U_{00}$, $U_{01}$, and $U_{11}$ have the form of (\ref{contact interaction}) with three s-wave scattering lengths characterizing the three independent types of binary collisions:
\begin{eqnarray}
U_{00} = \frac{4 \pi \hbar^2 a_{00}}{m} , \ \ 
U_{01} = \frac{4 \pi \hbar^2 a_{01}}{m} , \ \ 
U_{11} = \frac{4 \pi \hbar^2 a_{11}}{m} .
\label{U definitions}
\end{eqnarray}
The last term in 
(\ref{two boson hamiltonian}) describes resonant microwave-driven transitions between $ \KET{\Psi_0 } $ and $ \KET{\Psi_1 } $ with frequency $\Omega$.

The atoms are condensed into 
Gaussian translational modes 
$f_{0}({\bf r})$ and $ f_{1}({\bf r})$:
\begin{eqnarray}
\phi_{\alpha}({\bf r}) = f_{\alpha}({\bf r}) \, a_{\alpha}, \ \ 
f_{\alpha}({\bf r}) = \frac{ e^{ -r^2 / 2 \ell_{\alpha}^2} }{( \pi \ell_\alpha^2 )^\frac{3}{4} } , \ \ 
 \ell_{\alpha} = \sqrt{ \frac{\hbar }{m \omega_{\alpha}} }  , \ \ \alpha \in \{ 0,1\},
 \label{gaussian translational modes} 
\end{eqnarray}
where $ \int  \! d^3r  \, f_{\alpha} f_{\beta}  = 
[  \frac{2 \ell_\alpha \ell_\beta}{\ell_\alpha^2 + \ell_\beta^2 } ]^\frac{3}{2}$
and
$ \int  \! d^3r  \, f_{\alpha}^2 f_{\beta}^2  =  [ \frac{1}{ \pi (\ell_\alpha^2 + \ell_\beta^2) } ]^\frac{3}{2}$
(spatial modes overlap).
Here $a_\alpha^\dagger$ creates an atom with internal state $\KET{\Psi_\alpha}$ in a harmonic oscillator ground state $f_{\alpha}({\bf r})$ satisfying
$T_\alpha f_{\alpha}({\bf r})= \frac{\omega_\alpha}{2} f_{\alpha}({\bf r}) $,
and $[ a_\alpha , a^\dagger_{\beta} ] = \delta_{\alpha \beta} $.
Then
\begin{eqnarray}
H  =  \sum_{\alpha=0,1} \bigg( \! \frac{  \omega_\alpha}{2}  \, a_\alpha^\dagger a_\alpha   +   \frac{ u_{\alpha \alpha}}{2}  \, a_\alpha^\dagger 
a_\alpha^\dagger a_\alpha  a_\alpha  \! \bigg)   
+ u_{01} 
a_{0}^\dagger  a_{1}^\dagger a_0 a_1 + \lambda
 ( a_{0}^\dagger a_{1}  + a_{1}^\dagger a_{0}  ),
  \label{two-mode hamiltonian}
\end{eqnarray}
where 
\begin{eqnarray}
u_{00 }  = \frac{U_{00} }{  ( 2 \pi  \ell^2_{0} )^\frac{3}{2} } , \ \  
u_{01} 
 = \frac{U_{01}}{ [ \pi (\ell_0^2 \! + \! \ell_1^2) ]^\frac{3}{2}   } , \ \ 
 u_{11}  = \frac{U_{11} }{  ( 2 \pi  \ell^2_{1} )^\frac{3}{2} } , \ \ 
\lambda = \frac{\Omega}{2}  \bigg( \frac{ 2 \ell_0 \ell_1 }{\ell_0^2  \!  +  \!  \ell_1^2 } \bigg)^{ \! \frac{3}{2}} \! \! .
\label{u definitions}
\end{eqnarray}

We will use the two-mode model (\ref{two-mode hamiltonian}) as the basis for our investigations.
In the remainder of this section, we transform to the angular momentum representation and analyze squeezing. The spin-$\frac{\NUMBER}{2}$ Schwinger boson representation of SU(2) 
generators is given by
\begin{eqnarray}
& {\displaystyle{ J_x  =   \frac{ a_0^\dagger a_1 + a_1^\dagger a_0 }{2} , \ \ 
J_y = i \frac{  a_1^\dagger a_0 - a_0^\dagger a_1 }{2} , \ \ 
J_z =   \frac{a_0^\dagger a_0 -  a_1^\dagger a_1 }{2}}} ,
\ \ \NUMBER = a_0^\dagger a_0 + a_1^\dagger a_1 .
\label{schwinger boson}
\end{eqnarray}
These satisfy
\begin{eqnarray}
& J_{+} = J_{x} + i J_{y} = a_0^\dagger a_1, \ \ J_{-} = J_{x} - i J_{y}  = a_1^\dagger a_0 = J_{+}^\dagger , & \\
& [J^\mu , J^\nu ] = i  \epsilon^{\mu \nu \lambda} J^\lambda , \ \ 
[J_z, J_{\pm} ] = \pm  J_{\pm}, \ \ 
[J_{+} , J_{-} ] = 2  J_{z}. & 
\end{eqnarray}
To analyze squeezing it is sufficient to consider 
$\Omega = 0$ as well as  $\NUMBER \gg 1$. In this case  the two-mode Hamiltonian becomes
\begin{eqnarray}
H = \bigg(  \!  \frac{\omega_0 + \omega_1}{4} \! 
\bigg) \NUMBER
+ \bigg( \!  \frac{ u_{00}  + u_{11} + 2 u_{01} }{8} \! 
\bigg) \NUMBER^2
+ \bigg( \!  \frac{\omega_0 \!  - \!  \omega_1}{2} \! 
\bigg) J_z 
+ \bigg(  \frac{ u_{00} - u_{11} }{2} \bigg)  \NUMBER J_z 
+ \chi  J_z^2 , \ \ \ \ 
\end{eqnarray}
where 
\begin{eqnarray}
\chi := \frac{ u_{00} + u_{11} }{2}  - u_{01}  .
\label{chi definition}
\end{eqnarray}
Setting $ \omega_0 = \omega_1 $,  
$ u_{00} = u_{11}  $,
and dropping additive constants leads to the Kitagawa-Ueda one-axis twist model \cite{KitagawaUedaPRA1993}:
\begin{eqnarray}
H_{\rm KU} = \chi J_z^2.
\label{ku model}
\end{eqnarray}
Prepare a coherent state (\ref{coherent state}) centered at the $\KET{+}$ state on the Bloch sphere:
\begin{eqnarray}
\KET{ F_{+}  } =
 \frac{ ( a_0^\dagger + a_1^\dagger)^\NUMBER }{ \sqrt{2^\NUMBER} 
 {\sqrt{ \NUMBER !} } } \KET{{\rm vac}} .
\label{initial plus state}
\end{eqnarray}
The angular momentum
in the initial state (\ref{initial plus state}) is
\begin{eqnarray}
 \langle J_{x} \rangle = \frac{\NUMBER}{2}, \ \ \langle J_{y} \rangle  = \langle J_{z} \rangle = 0, \ \ 
 \langle J_{x}^2 \rangle  = \frac{\NUMBER^2}{4} , \ \ 
 \langle J_{y}^2 \rangle  =  \langle J_{z}^2 \rangle  = \frac{\NUMBER}{4} ,
\end{eqnarray}
and the variances are
$ {\rm Var}(J_x)  = 0$
and 
${\rm Var}(J_y) = {\rm Var}(J_z)  = \frac{\NUMBER}{4} .$
We note that $\KET{ F_{+}  }$  is a minimum-uncertainty state satisfying
\begin{eqnarray}
 {\rm Var}(J_y) =  {\rm Var}(J_z) =
\frac{ |  \langle  [J_y , J_z] \rangle | }{2} =
\frac{  \langle J_x \rangle  }{2}
= \frac{\NUMBER}{4} . 
\end{eqnarray}

Next turn on the atomic interaction (with $ \omega_0 = \omega_1 $ and $ u_{00} = u_{11} $). This leads, in the spin picture, to the Kitagawa-Ueda Hamiltonian (\ref{ku model}).
After evolution for a time $t$ the state becomes
$  U \KET{ F_{+} } $, where $U=e^{-i \chi  t J_z^2}$. 
The squeezing occurs in the $yz$ plane, so  define
\begin{eqnarray}
J_{\! \varphi} := \sin(\varphi) J_y + \cos(\varphi)  J_z ,
\end{eqnarray}
and calculate the expectation and variance of $ J_{\! \varphi} $ as a function of angle $ \varphi $.  Use the identity
\begin{eqnarray}
U^\dagger J_{+} U = J_{+}  \, e^{2 i  \chi t  J_z} e^{i \chi t } , \ \ 
U=e^{-i \chi  t J_z^2},
\label{jplus identity}
\end{eqnarray}
to obtain 
\begin{eqnarray}
 \langle J_{\! \varphi} \rangle =  \BRA{F_{+}} U^\dagger J_{\! \varphi} U \KET{F_{+}} = 0  , \ \ t \ge 0,  \end{eqnarray}
and \cite{WangMolmerEPJD2002} 
\begin{eqnarray}
 {\rm Var}(J_{\! \varphi} ) 
 =  \frac{\NUMBER}{4}  + \frac{\NUMBER^2}{8}  
 \bigg[1 \! - \! \cos^{\NUMBER-2}(2 \chi t ) \bigg]\sin^2(\varphi)
+  \frac{\NUMBER^2}{4}   \cos^{\NUMBER-2}(\chi t)  
\sin(\chi t)
\sin(2\varphi) ,
\end{eqnarray}
where we have used $\NUMBER \gg 1$.
We conclude  that
the two-mode model (\ref{two-mode hamiltonian}) correctly recovers established properties of spin squeezed condensates \cite{\SPINSQUEEZING}. 

\subsection{Path integral and large $\NUMBER$ limit}
\label{path integral and large n limit}

Next we take the $\NUMBER \rightarrow \infty $ limit and explain how to perform single-input QSD with the same Hamiltonian
(\ref{two-mode hamiltonian}). This is an adaptation of \cite{240416288} to the 
two-component BEC. To use the BEC as a nonlinear qubit, we need to control the accuracy of mean field theory by making $\NUMBER$ large and leveraging entanglement monogamy. This keeps the many-body wave function close to a coherent state (\ref{coherent state}) for some qubit coordinates (probability amplitudes) $\psi_{0,1} \in {\mathbb C}$, even when the interaction is present (without loss of generality we assume here that $\psi_0$ is real and nonnegative). The only relevant dynamical variables in this limit are the coordinates themselves, and we are interested in their propagator $\BRA{F } e^{-i H t  } \KET{ F^\prime }$.  Here $\KET{ F}$ and $\KET{ F^\prime }$ are states of the form (\ref{coherent state}), with qubit coordinates   $\psi_{0,1}$ and  $\psi^\prime_{0,1}$ respectively.

In some cases it is possible to calculate the  dynamics exactly for any $\NUMBER$ (Section \ref{kitagawa-ueda model large n}). More generally we seek a simplified description of the problem, made possible by letting  $\NUMBER \rightarrow \infty $. Note that 
$ \langle F | F^\prime \rangle =
\langle \psi | \psi^\prime \rangle^{\NUMBER} $,
where $\langle \psi | \psi^\prime \rangle = {\bar \psi}_{0} \psi^\prime_{0} + {\bar \psi}_{1} \psi^\prime_{1}$ is the single-atom overlap (and
$| \langle \psi | \psi^\prime \rangle| \le 1$).
Here ${\bar z }$ denotes complex conjugation. 
It follows that, for any $ \KET{F} \neq \KET{F^\prime} $, we must have $ \lim_{\NUMBER \rightarrow \infty} \langle F | F^\prime \rangle  = 0$. 
This orthogonality catastrophe defines a quasiclassical limit for the model
(where coherent states are mutually orthogonal),  distinct from the $\hbar \rightarrow 0$ limit. The coherent states occupy a subspace of two-mode Fock space that we call ${\cal H}_{\rm coh}$. Let
\begin{eqnarray}
\KET{n_0 ,n_1} := 
\frac{ (a_0^\dagger)^{n_0} }{\sqrt{ n_0 ! \, } } 
\frac{ (a_1^\dagger)^{n_1} }{\sqrt{ n_1 ! \, } } 
 \KET{{\rm vac}} , \ \ n_0, n_1 \in \{ 0,1,2, \cdots \}.
\end{eqnarray}
 Expanding (\ref{coherent state}) in this basis leads to
$\KET{ F} 
 = \sum_{k=0}^\NUMBER
\sqrt{   \binom{\NUMBER}{k}  } \, \psi_0^{\NUMBER-k} \psi_1^{k}  \,\KET{\NUMBER \! - \! k ,k} $,
where $\binom{\NUMBER}{k}$ is the binomial coefficient. Therefore ${\rm dim}( {\cal H}_{\rm coh}) = \NUMBER + 1$.
Averaging the pure state $\KETBRA{F}$ over the Bloch sphere yields a maximally mixed state
$ \int \! \frac{ \sin(\theta) d\theta d\phi }{4 \pi}
 \KETBRA{F} \! = \! \frac{ I_{\NUMBER \! + \! 1} }{\NUMBER \! + \! 1}$, 
 where $I_{\NUMBER \! + \! 1} = \sum_{k=0}^\NUMBER \KETBRA{\NUMBER \! - \! k,k} $ is the identity in ${\cal H}_{\rm coh}$. 
 
 Now split the evolution into 
$K \! \gg  \! 1$ short-time  propagators. This leads to
$ \BRA{F } e^{-i H t  } \KET{ F^\prime } =
\BRA{F } (e^{-i H \tau })^K \KET{ F^\prime } 
=  \int ( \prod_{k=1}^{K-1} \! \frac{\NUMBER \! + \!  1}{4 \pi} \sin \theta_k d\theta_k d\phi_k ) \prod_{k=1}^{K} 
\BRA{F_{k} } e^{-i H \tau  } \KET{ F_{k-1} }$, 
where $\KET{ F_0 } = \KET{ F^\prime}$,
$\KET{ F_K } = \KET{ F}$, and
$\tau = t/K$.
For small 
$\tau$ and large $\NUMBER$ we have 
\begin{eqnarray}
\BRA{F } e^{- i H \tau}  \KET{ F^\prime  } 
& = & \langle F | F^\prime  \rangle  
\bigg\{ 1 -  i \NUMBER  \tau \!  \bigg[  \! 
 \sum_{\alpha} \! \bigg( \! \frac{  \omega_{\alpha} }{2}  \,  {\bar \psi}_{\alpha} \psi_{\alpha}   +   \frac{u_{\alpha \alpha}}{2}  \NUMBER  \,
  {\bar \psi}_{\alpha}  {\bar \psi}_{\alpha}
{\psi}_{\alpha} {\psi}_{\alpha}  \! \bigg)  \nonumber \\
 &+&  u_{01} \NUMBER  {\bar \psi}_{0}  {\bar \psi}_{1} 
{\psi}_{0} {\psi}_{1} 
+ \lambda (  {\bar \psi}_{0} {\psi}_{1} 
 +  {\bar \psi}_{1} {\psi}_{0} )  \bigg]  
 + O(\tau^2) \bigg\} .
\end{eqnarray}
The orthogonality catastrophe implies that
$\langle F | F^\prime  \rangle$ vanishs unless  $\KET{F} 
\approx \KET{F^\prime} $, allowing us to set $\psi_{\alpha}^\prime \! = \! \psi_{\alpha} $ in the $O(\tau)$ term. 
Let 
\begin{eqnarray}
\int \!  D {\bar \psi}_{\alpha} \, D \psi_{\alpha}   \,  \delta( | \psi_0 |^2 \! + \!  | \psi_1 |^2 \! - \!  1) 
 :=  \lim_{K \rightarrow \infty} \int  (  \! 
 \prod_{k=1}^{K-1} \! \frac{\NUMBER \! + \!  1}{4 \pi} \sin \theta_k d\theta_k d\phi_k  \! ) 
 \end{eqnarray}
 denote a sum over Bloch sphere paths 
 $\psi_{\alpha}(t_k)$ connecting $ \KET{F^\prime} $ to $ \KET{F} $, with $ t_k = k \tau $ . Then we have
 $ \langle F_{k} | F_{k-1} \rangle = e^{-\NUMBER \tau \sum_\alpha {\bar \psi}_\alpha \partial_t \psi_\alpha + O(\tau^2)} $ and the propagator can be written in a spin-coherent-state path integral representation as 
\cite{HaldanePRL1988,LossDiVincenzoPRL1992}
\begin{eqnarray}
 \BRA{F } e^{-i H t  } \KET{ F^\prime } 
 = \int \!  \! D {\bar \psi}_{\alpha} \, D \psi_{\alpha}   \,  \delta( | \psi_0 |^2 \! + \!  | \psi_1 |^2 \! - \!  1) \,
 e^{ i  S } , \ \ 
 S = \int \! dt  \,
\BRA{ F }  i  \partial_t - H
\KET{F } , \ \ \ \ 
 \end{eqnarray}
where
\begin{eqnarray}
 S =   \NUMBER  \! \!  \int \!  \! dt 
\bigg\lbrace \! 
 \sum_{ \alpha =0,1} \! \bigg( \!  {\bar \psi}_\alpha   i \partial_t  \psi_\alpha
- \frac{  \omega_{\alpha} }{2}  \, 
| \psi_{\alpha} |^2  
-  \frac{ u_{\alpha \alpha}}{2}
 \NUMBER  \,  |{\psi}_{\alpha} |^4 \! \bigg)
- u_{01} \NUMBER | \psi_0  \psi_1 |^2
- \lambda (  {\bar \psi}_{0} {\psi}_{1} 
 +  {\bar \psi}_{1} {\psi}_{0} ) \! \bigg\rbrace . \ \ \ \ \ 
\label{effective action}
\end{eqnarray}

As we have explained in Section \ref{large n limit and 2-particle entanglement}, to take the large $\NUMBER$ limit of the condensate model
we must assume that the limits
 \begin{eqnarray}
K_{00} = \lim_{\NUMBER \rightarrow \infty} \NUMBER u_{00} , \ \ 
K_{01} = \lim_{\NUMBER \rightarrow \infty} \NUMBER u_{01}, \ \
{\rm and}  \ \
K_{11} = \lim_{\NUMBER \rightarrow \infty} \NUMBER u_{11},
\label{k definitions}
 \end{eqnarray}
exist. Then in the large $\NUMBER$ limit the stationary phase equations are
\begin{eqnarray}
&& i \frac{d \psi_{0} }{dt} = \bigg( \! \frac{\omega_0}{2} + K_{00}  |\psi_0|^2 + K_{01}   |\psi_1|^2 \bigg) \psi_0
 + \lambda \, \psi_1 , \\
 && i \frac{d \psi_{1} }{dt} = \bigg( \!  \frac{\omega_1}{2} + K_{11}   |\psi_1|^2 + K_{01}   |\psi_0|^2 \bigg) \psi_1
 + \lambda \, \psi_0 ,
  \label{mean field equations separate}
 \end{eqnarray}
 as well as their complex conjugates. Therefore we obtain
\begin{eqnarray}
 \frac{d}{dt} 
 \begin{pmatrix}
\psi_0   \\  \psi_1 
\end{pmatrix} 
= -i H_{\rm eff} \!
\begin{pmatrix}
\psi_0   \\  \psi_1 
\end{pmatrix} \! ,
 \label{mean field equations combined}
 \end{eqnarray}
where  
\begin{eqnarray}
 H_{\rm eff} &=& 
\begin{pmatrix}
\frac{\omega_0}{2} + K_{00} |\psi_0|^2 + K_{01} |\psi_1|^2 & \lambda  \\  \lambda &
 \frac{\omega_1}{2} + K_{11} |\psi_1|^2 + K_{01} |\psi_0|^2 
\end{pmatrix} \\
&=& \bigg[ \frac{\omega_0 + \omega_1}{4}
+ \frac{ K_{00} + K_{11} + 2 K_{01}}{4}
+ (|\psi_0|^2-|\psi_1|^2) 
\frac{ K_{00} - K_{11} }{4} 
\bigg]  \!
\begin{pmatrix}
1 & 0 \\ 0 & 1
\end{pmatrix} \\
&+ &  B_x \!
\begin{pmatrix} 0 & 1 \\ 1 & 0 \end{pmatrix}  
+  B_z \! \begin{pmatrix}
1 & 0 \\ 0 & -1
\end{pmatrix} + g  (|\psi_0|^2-|\psi_1|^2) \\
&= &  {\rm constant} + B_x \sigma^x + B_z \sigma^z 
+ g \,  {\rm tr}(\rho \sigma^z) \sigma^z ,
\label{heff original}
\end{eqnarray}
with
\begin{eqnarray}
B_x := \lambda, \ \ 
B_z :=  \frac{\omega_0 - \omega_1}{4}
+\frac{ K_{00} - K_{11} }{4}  , \ \ 
 g := \frac{ K_{00} + K_{11} -  2 K_{01} }{4} .
 \label{b and g}
 \end{eqnarray}
 This expression for $H_{\rm eff}$ is the main result of this section, but the
total energy also includes a ``background'' energy term common to mean field theory:
\begin{eqnarray}
H_{\rm tot} = H_{\rm eff} +  \Delta E, \ \ 
 \Delta E = - \frac{g}{2} \, [ {\rm tr}(\rho \sigma^z) ]^2 .
 \label{background}
 \end{eqnarray}
For example, consider 
a spin model
$ H_{\rm spin} = \frac{g}{\NUMBER} \sum_{1 \le i < j \le \NUMBER} \sigma_i^z \sigma_j^z $ with 
all-to-all interaction.
Expanding $H_{\rm spin} = H_{\rm lin} 
+ H_{\rm quad}$ in powers of spin fluctuations $\sigma_i^z - \langle \sigma^z  \rangle$
and assuming a homogeneous order parameter
$\langle \sigma^z  \rangle $, and $\NUMBER \gg 1$, leads to 
$ H_{\rm lin}  = g
\sum_{i=1}^\NUMBER [  \langle\sigma^z\rangle \sigma_i^z  - \frac{1}{2} \langle\sigma^z\rangle^2 ] .
$ In the mean field approximation $H_{\rm quad} $ is neglected and
\begin{eqnarray}
H_{\rm spin}  \approx  \NUMBER
[ g \langle\sigma^z\rangle \sigma^z  \! - \! \frac{g}{2}  \langle\sigma^z\rangle^2
]  = \NUMBER
[ H_{\rm eff}  - \! \frac{g}{2} \langle\sigma^z\rangle^2
]  .
\end{eqnarray}

\clearpage

\section{Single-input state discrimination}
\label{single-input state discrimination section}

\subsection{Torsion and conservation laws}
  
To implement fast state discrimination, we consider the symmetric setting 
$\omega_{0} = \omega_{1} =  \omega$ 
and $K_{00} = K_{11} = K$ in (\ref{heff original}), which leads to the torsion model (\ref{torsion hamiltonian}) plus a $\sigma^x $ term: 
 \begin{eqnarray}
 H_{\rm eff}  = B_x \sigma^x 
+  g (|\psi_0|^2 \! - \!  |\psi_1|^2)  \sigma^z  , \ \ 
g = \frac{K - K_{01}}{2} .
 \label{heff}
 \end{eqnarray}
 The same result is obtained in the absence of that  symmetry after adding another $\sigma^z$ operator to  the Hamiltonian to cancel the nonvanishing $B_z \sigma^z$ term in (\ref{heff original}). 
In the Pauli basis 
 $ r^\mu = {\rm tr}(\rho \sigma^\mu)$ 
we have
 $ \frac{d r^\mu}{dt}  = -i \, {\rm tr}( \!  \rho [ \sigma^\mu , H_{\rm eff} ] )$ 
and the torsion model (\ref{heff}) takes the form
 \begin{eqnarray}
&& \frac{dx}{dt} = - 2 g y z , 
\label{torsion model x} \\
&& \frac{dy}{dt} = 2 gxz-2 B_x z ,
\label{torsion model y} \\
&& \frac{dz}{dt} = 2 B_x y .
\label{torsion model z} 
\end{eqnarray}
These equations will be used to design a
single-input QSD gate. The solutions to these equations when $B_x = 0$ are shown in Figure~\ref{z torsion figure}.

We now consider a single-input QSD gate that
accepts one of two states 
$\{ \KET{a} , \KET{b} \}$
as the input, and maps  $\KET{a} \mapsto \KET{0}$ and $\KET{b} \mapsto  \KET{1}$. 
This operation is impossible within linear one-particle quantum mechanics
(unless $ \langle a | b \rangle = 0$)  because
it increases the trace distance. Note, however,  that this QSD gate is distinct from 
(\ref{discrimination problem for 3sat}), which must 
map a large set of potential inputs to $\KET{1}$.

We are faced with a control problem on the Bloch sphere. A unitary gate can always be used to orient the pair $\{  {\vec r}_{\rm a},  {\vec r}_{\rm b} \} $  anywhere on the Bloch sphere without changing the angle between them. Childs and Young \cite{150706334} constructed a gate by maximally increasing the trace distance
$  \| \rho_{\rm a}  -  \rho_{\rm b} \|_1
= | {\vec r}_{\rm a} - {\vec r}_{\rm b} | $ until
$ | {{\vec r}_{\! \rm a}}^{\, \prime} -  {{\vec r}_{\! \rm b}}^{\, \prime} | =2 $
(the Bloch vectors become antiparallel), after which a  unitary 
$ U_{\rm read} = \KET{0} \BRA{a^\prime} + \KET{1} \BRA{b^\prime} $ is applied prior to measurement in the $\{ \KET{0} , \KET{1} \}$ basis. This leads to a fast gate but requires a time-dependent $B_x$.
Here we take a different approach using a time-independent $B_x$ and no readout unitary.
The Bloch coordinates in the torsion model 
(\ref{torsion model x}) - (\ref{torsion model z})  satisfy two conservation laws. The first is conservation of total energy (\ref{background}): 
\begin{eqnarray}
E = B_x x + \frac{g}{2} z^2,  \ \ 
\frac{dE}{dt} = B_x \frac{dx}{dt} + g z \frac{dz}{dt} = 0.
 \end{eqnarray}
 Here we have assumed that $B_x$ and $g$ are constant. The second is conservation of the Bloch vector length (hence purity and entropy):
\begin{eqnarray}
r^2 = x^2 + y^2 + z^2 ,  \ \ 
\frac{dr^2}{dt} = 
2  x \frac{dx}{dt} + 2  y \frac{dy}{dt} + 2  z \frac{dz}{dt} = 0.
 \end{eqnarray}
    
\begin{figure}
\begin{center}
\includegraphics[width=8.0cm]{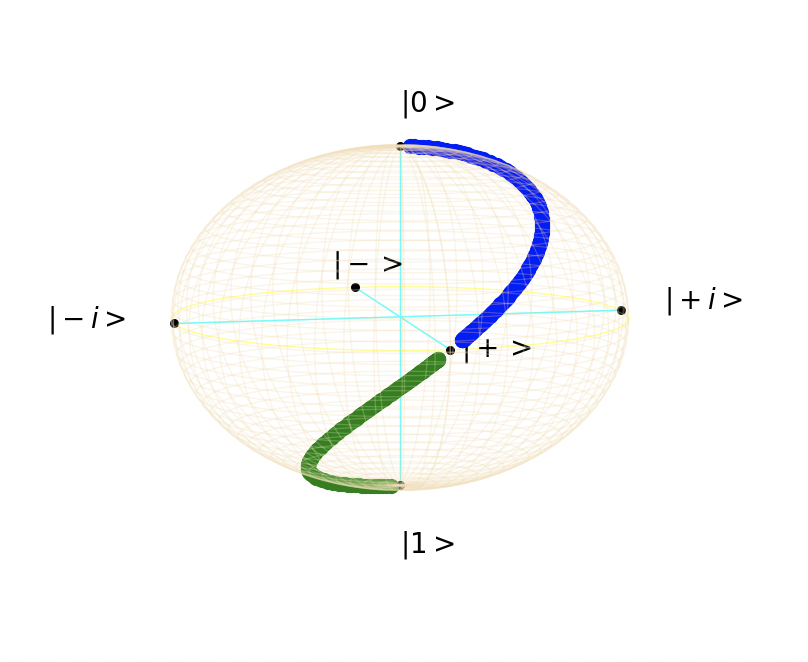} 
\caption{Viviani curve mapping $\KET{a} \mapsto \KET{0}$ (blue) and $\KET{b} \mapsto  \KET{1}$ (green).} 
\label{viviani figure}
\end{center}
\end{figure} 

\subsection{State discrimination along Viviani's curve}

Assume that $g > 0$. Setting $E = B_x = \frac{g}{2}$ leads to striking orbits in the shape of Viviani's curve, shown in Figure~\ref{viviani figure}, at the intersection of a  cylinder (oriented in the $z$ direction) and the Bloch sphere. 
To understand why, note that when $E = B_x = \frac{g}{2}$ the energy conservation equation for a pure state ($r = 1$) is
 $x + z^2 = 1 = x^2 + y^2 + z^2$,
or
$ x^2 - x + y^2 = 0$, 
which is the equation for a cylinder of radius 
 $\frac{1}{2}$ tangent to the Bloch sphere at the 
point ${\vec r} = (1,0,0)$:
\begin{eqnarray}
(x-\textstyle{\frac{1}{2}})^2 + y^2 = (\textstyle{\frac{1}{2}})^2 .
\end{eqnarray}
Thus we seek an orbit in the shape of  
Viviani's curve:
\begin{eqnarray}
x = \cos^{2} \VIVIANI , \ \ 
y = \cos \VIVIANI \, \sin \VIVIANI, \ \ 
z  = \sin \VIVIANI  , \ \  
- \frac{\pi}{2} \le \VIVIANI \le \frac{\pi}{2}.
\label{viviani curve}
\end{eqnarray}
Note that (\ref{viviani curve}) satisfies (\ref{torsion model z}) if the state traverses the curve with $\VIVIANI$ evolving as
\begin{eqnarray}
\VIVIANI(t) = 2 \arctan  \! \bigg[ \! \tan 
\bigg(  \!  \frac{\VIVIANI(0)}{2}  \!  \bigg)
 e^{2B_xt}  \bigg] , \ \ 
 \frac{d\VIVIANI}{dt} = 2B_x \sin \VIVIANI.
\end{eqnarray}
Also note that (\ref{viviani curve}) satisfies (\ref{torsion model x}) 
and (\ref{torsion model y}) if $B_x = \frac{g}{2}$. 
As in the approach of Childs and Young \cite{150706334}, qubit states $\{ \KET{a} , \KET{b} \}$ are initialized to
Bloch coordinates
\begin{eqnarray}
 x_{a} = x_{b} 
= \bigg|  \! \cos\bigg( \! \frac{  \theta_{\rm ab}} {2}  \! \bigg) \bigg|, \ \ 
y_{a}= z_{a} = \frac{ \sin ( \frac{  \theta_{\rm ab}} {2})}{\sqrt 2}, \ \  
y_{b}= z_{b} = - \frac{ \sin ( \frac{ \theta_{\rm ab}} {2})}{\sqrt 2} ,
\ \ 0 \le \theta_{\rm ab} \le \pi,
\label{initial viviani states}
\end{eqnarray}
where $ \theta_{\rm ab} $ is the  initial angle between the Bloch vectors ${\vec r}_{\rm a}$ and  ${\vec r}_{\rm b}$.  For small initial $\theta_{\rm ab}$, the Bloch vectors ${\vec r}_{\rm a,b}$ lie on Viviani's curve with  $ \VIVIANI(0) =  \pm (\frac{ \theta_{\rm ab}} {2 \sqrt{2}} ) $, where the $+(-)$ sign applies to  ${\vec r}_{\rm a} \, ({\vec r}_{\rm b})$. Subsequent evolution of ${\vec r}_{\rm a}$ follows a Viviani curve segment  ending at classical state 
$ \KET{0} $ where
$\VIVIANI = \frac{\pi}{2}$ (the blue curve in Figure~\ref{viviani figure}). This occurs after a time 
\begin{eqnarray}
 t_{\rm V}  = \frac{1}{g}  \log \cot \bigg( \! \frac{ \theta_{\rm ab}} {4 \sqrt{2}}  \!  \bigg) \approx \frac{1}{g} \log \bigg( \! \frac{4 \sqrt{2} }{\theta_{\rm ab} }  \! \bigg),
 \label{tv}
\end{eqnarray}
where the second expression assumes
$\theta_{\rm ab} \ll 1$. 
Similarly, ${\vec r}_{\rm b}$ follows a Viviani curve segment  down to $ \KET{1} $ at 
$\VIVIANI = -\frac{\pi}{2}$
 (the green curve).
If  $g<0$, the
initial states
 (\ref{initial viviani states}),
 after reflection though the $y \! = \! 0$ plane, follow other Viviani curve segments to the poles.
The Viviani gate is efficient in the sense that an exponentially small $\theta_{\rm ab} = 2^{-k}$ 
takes time  $t_{\rm V} = O(k)/g$.

\clearpage

\subsection{Kitagawa-Ueda model as $\NUMBER \rightarrow \infty$}

\label{kitagawa-ueda model large n}

Bloch sphere torsion provides a quantum information processing advantage by circumventing trace distance monotinicity
(\ref{trace distance monotinicity}) and enabling 
expansive qubit dynamics. 
How do we know this is a real effect and not an artifact of the stationary phase approximation or mean field theory? In the remainder of this section we show that the dynamics of pure torsion ($B_x \!  = \! 0$) shown in Figure~\ref{z torsion figure} follows directly from the exactly solvable Kitagawa-Ueda spin model (\ref{ku model})
in the double limit $\NUMBER \rightarrow \infty, \ \chi \rightarrow 0$. In this model the angular momentum is carried by $\NUMBER$ 
pseudospins 
\begin{eqnarray}
J_{\mu} = \sum_{i=1}^\NUMBER \frac{\sigma_i^\mu}{2} , \ \ \mu \in \{x,y,z \} ,
\label{angular momentum from spins}
\end{eqnarray}
instead of the boson fields as in (\ref{schwinger boson}). It is  possible to explicitly calculate 
$\langle J_{\mu} \rangle $ as a function of time in the Kitagawa-Ueda model, starting in any $\NUMBER$-qubit permutation-symmetric product 
state $ \KET{\Psi_{\NUMBER}} =  \KET{\psi}^{\! \otimes \NUMBER} \! $, where $\KET{\psi}$ is a single-qubit state. From (\ref{jplus identity}) we obtain
\begin{eqnarray}
 \langle J_{+} \rangle = \BRA{\Psi_{\NUMBER}} \, U^\dagger (J_{x} + i J_{y} ) U 
\KET{\Psi_{\NUMBER}}
= \BRA{\Psi_{\NUMBER}} J_{+}   
e^{2 i \chi t J_z} e^{i \chi t} 
\KET{\Psi_{\NUMBER}}  , \ \ U = e^{-i \chi t J_z^2}.
\end{eqnarray}
Using (\ref{angular momentum from spins}) then leads to
\begin{eqnarray}
 \langle J_{+} \rangle &=&  \frac{\NUMBER}{2} e^{i \chi t} 
 \BRA{\psi} (\sigma^{x}  + i \sigma^{y} ) 
 e^{i \chi t \sigma^{z}  }   \KET{\psi} \,
  \big[ \BRA{\psi}  e^{i \chi t \sigma^{z}  }   \KET{\psi} \big]^{\NUMBER-1} \\
  &=& \frac{\NUMBER}{2} (x+iy) 
\big[  \cos(\chi t) + i z  \sin(\chi t) 
\big]^{\NUMBER-1} .
 \end{eqnarray}
Here $x, y, z$ are the Bloch coordinates of the {\it initial} single-qubit state $\KET{\psi}$. 
In addition, 
\begin{eqnarray}
 \langle   J_{z} \rangle = \frac{\NUMBER}{2} z ,
\end{eqnarray}
because $J_{z} $ commutes with the Hamiltonian
(\ref{ku model}). 

Next we take the double limit $ \NUMBER \rightarrow \infty, \ \chi \rightarrow 2g/\NUMBER $, 
with $g$ constant [the factor of two here comes from  (\ref{chi definition}) and (\ref{b and g})].
The technical part of the calculation is to evaluate 
\begin{eqnarray}
f(t) := \lim_{\NUMBER \rightarrow \infty} 
  \bigg[ \! \cos\bigg( \! \frac{2gt}{\NUMBER} \bigg)  + i z  \sin\bigg( \! \frac{2gt}{\NUMBER} \bigg) 
    \bigg]^{\NUMBER-1}  \! \! .
 \label{ku f function}
\end{eqnarray}
To do this, rewrite it as  
$f(t) = \lim_{\NUMBER \rightarrow \infty}
[  c_{\NUMBER}(t)]^{\NUMBER-1}$, where
$ c_{\NUMBER}(t) = \cos( \! \frac{2gt}{\NUMBER})  + i z \sin( \! \frac{2gt}{\NUMBER})$.
Then expand $f(t)$ as a Maclaurin series in time,
\begin{eqnarray}
 f(t) = f(0) + \sum_{k=1}^\infty  \frac{t^k}{k!} \, f^{(k)}(0) ,
\end{eqnarray}
 with
 $ f^{(k)}(t) := \frac{d^k \! f}{dt^k} $.
 Note that
\begin{eqnarray}
 \frac{df}{dt} = \lim_{\NUMBER \rightarrow \infty} 
\, (\NUMBER \! - \! 1) 
(  c_{\NUMBER} )^{\NUMBER-2} \,\frac{dc_{\NUMBER}}{dt} , \ \ 
\frac{dc_{\NUMBER}}{dt} = \bigg[ \! - \!  \sin\bigg( \! \frac{2gt}{\NUMBER} \bigg)  + i z \cos\bigg( \! \frac{2gt}{\NUMBER} \bigg) 
\bigg] \frac{2g}{\NUMBER},
\label{dfdt}
 \end{eqnarray}
 and
 \begin{eqnarray}
\frac{d^2 \! f}{dt^2} = \lim_{\NUMBER \rightarrow \infty} 
\bigg[  (\NUMBER \! - \! 1) (\NUMBER \! - \! 2) 
(  c_{\NUMBER} )^{\NUMBER-3} \bigg( \! \frac{dc_{\NUMBER}}{dt}  \! \bigg)^{\! 2}
+ \, (\NUMBER \! - \! 1) 
(  c_{\NUMBER} )^{\NUMBER-2} \,\frac{d^2c_{\NUMBER}}{dt^2} \bigg].
\label{d2fdt2}
 \end{eqnarray}
 
Because  $ \frac{d^k \! c_\NUMBER}{dt^k} = O(\NUMBER^{-k}) $, we see that
$ (\frac{dc_{\NUMBER}}{dt})^{2} $
and
$ \frac{d^2c_{\NUMBER}}{dt^2}  $
are of the same order, so the first term in 
(\ref{d2fdt2}) determines the large $\NUMBER$ limit. This behavior extends to the higher order derivatives:
$ \frac{d^k \! f}{dt^k} = \lim_{\NUMBER \rightarrow \infty} 
\NUMBER^k
(  c_{\NUMBER} )^{\NUMBER-k-1} (  \frac{dc_{\NUMBER}}{dt} )^{k}
$ and 
$ f^{(k)}(0) = \NUMBER^k ( \frac{ 2 i g z}{\NUMBER} )^k =  (2 i g z)^k $.
Therefore we obtain
\begin{eqnarray}
 f(t) = e^{2 i g z t} ,
\end{eqnarray}
leading to
\begin{eqnarray}
&& \lim_{\NUMBER \rightarrow \infty}
\frac{  \langle   J_{x} \rangle }{\NUMBER/2} = x
\cos(2gzt) - y \sin(2gzt) , \\
&& \lim_{\NUMBER \rightarrow \infty}
\frac{  \langle   J_{y} \rangle }{\NUMBER/2} = x
\sin(2gzt) + y \cos(2gzt) , \\
&& \lim_{\NUMBER \rightarrow \infty}
\frac{  \langle   J_{z} \rangle }{\NUMBER/2} = z .
\end{eqnarray}
To interpret this result, define a vector   
${\vec j} = (j_{x},j_{y},j_{z})$ from the 
normalized angular momentum components 
 $ j_\mu := \lim_{\NUMBER \rightarrow \infty}
\frac{  \langle   J_{\mu} \rangle }{\NUMBER/2}, \ \mu \in \{ x,y,x\} $, and
note that 
$ |{\vec j}(t) | =1$
and ${\vec j}(0) = (x,y,z)$,
where $x,y,z$ are the Bloch vector components of the initial qubit state. Then we find
\begin{equation}
\begin{pmatrix} j_{x} \\ j_{y} \\ j_{z}  \end{pmatrix}
=
\begin{pmatrix} 
\cos (2gzt) & -\sin(2gzt) & 0  \\
\sin(2gzt) & \cos(2gzt) & 0  \\ 
0  & 0  & 1  \\ 
\end{pmatrix}
\begin{pmatrix} j_{x} \\ j_{y} \\ j_{z} \end{pmatrix}_{\! \! t=0} \!  \! \! \! \! .
\end{equation}
The matrix implements $z$-axis torsion: A $z$ rotation by angle $2g z t$ proportional to $z$.

\section{Fast discrimination with torsion and dissipation}
\label{torsion with dissipation section}

\subsection{Combining contractive  and expansive dynamics}

The appeal of nonlinear 
QSD based on the Childs and Young or 
Viviani gate is that, assuming knowledge of $\KET{a}$ and 
$\KET{b}$, a protocol can be designed to implement perfect QSD without multiple copies of the input. However the input $ {\bf r} \in \{ {\bf r}_{\rm a} , {\bf r}_{\rm b} \} $ must be free of errors (in this section we generalize to mixed states described by Bloch vectors). In previous work \cite{211105977} we considered the combined effects of torsion and dissipation, and investigated a discriminator based on the generation of two basins of attraction 
$ {\mathbb V}_{\! +} , {\mathbb V}_{\! -}   \subset  {\mathbb B} $ 
within the Bloch ball $ {\mathbb B}  $, flowing to opposing stable fixed points denoted by Bloch vectors 
$ \{ {\bf r}^{\rm fp}_{+} , {\bf r}^{\rm fp}_{-} \}$. The nonoverlapping subsets $ {\mathbb V}_{\! \pm}  $
share a two-dimensional boundary cutting through the center of the Bloch ball, across which the discriminator inputs 
$ \{ {\bf r}_{\rm a} , {\bf r}_{\rm b} \}$ are placed. The  fixed points $ \{ {\bf r}^{\rm fp}_{+} , {\bf r}^{\rm fp}_{-} \}$ are the discriminator outputs. This dynamics leads to an intrinsic tolerance to errors in the input  
$ {\bf r} \in \{ {\bf r}_{\rm a} , {\bf r}_{\rm b} \} $ 
that do not cross the boundary into the wrong basin \cite{211105977}. 
The gate combines contractive dynamics within the basins of attraction $ {\mathbb V}_{\! \pm},$ made possible by the dissipation, with expansive dynamics at their boundary separatrix 
made possible by the torsion.

Here we develop this idea further, focusing 
on a related computational benefit of the 
nonlinear  dynamics: autonomy. By this we mean that the QSD gate and experimental protocol do not depend on the values of $ \{ {\bf r}_{\rm a} , {\bf r}_{\rm b} \}$, and their prior knowledge is no longer required. In particular, we design an {\it autonomous} QSD gate that distinguishes any state in subset $ {\mathbb V}_{\! +} $ from any in $ {\mathbb V}_{\! -} $, where  $ {\mathbb V}_{\! \pm} $  are the sets of inputs 
(basins of attraction)
feeding the stable fixed points 
$ {\bf r}^{\rm fp}_{\pm} $. 
Note that the autonomous QSD gate discussed here is no longer a single-input protocol because  the outputs $ {\bf r}^{\rm fp}_{\pm} $ are not perfectly distinguishable. However, because the fixed points $ {\bf r}^{\rm fp}_{\pm} $ are known explicitly (see below), we expect that a second step mapping them to the perfectly distinguishable classical states $ \{ \KET{0}, \KET{1} \}$ can be implemented, making the gate both single-input and autonomous (we do not discuss the second step here).
  
Let's put nonlinearity aside briefly and recall how dissipation and decoherence are included in the  Bloch sphere picture
\begin{eqnarray}
r^\mu  = {\rm tr}(\rho \sigma^\mu ) , \ \ 
\frac{dr^\mu}{dt}  = {\rm tr} \bigg( \! \frac{d\rho}{dt} \sigma^\mu \!\bigg) , 
\label{bloch sphere picture}
\end{eqnarray}
where $\mu \in \{ x,y,z\}$.
The most general linear time-local equation of motion (linear Markovian master equation) for a quantum state 
$\rho \in \CMATRIX{d} $ is
\begin{eqnarray}
\frac{d \rho}{dt} = -i [H , \rho ]  + \sum_\alpha
\zeta_\alpha \, B_\alpha \rho B_\alpha^\dagger
+ \{ L_{+} , \rho \}, \ \ L_{+} := - \frac{1}{2} \sum_\alpha \zeta_\alpha  B_\alpha^\dagger B_\alpha .
\label{master equation}
\end{eqnarray}
Here $H  \in \CMATRIX{d} $ is a Hermitian operator 
generating unitary evolution, the  
$ B_\alpha \in \CMATRIX{d} $ are linear but otherwise arbitrary jump operators generating
dissipative and decohering processes, 
and the $ \{ L_{+} , \rho \} $ term conserves the trace of $\rho$.

The arbitrary signs $ \zeta_\alpha \in \pm 1$ determine the degree of positivity preserved by the dynamics: If all $ \zeta_\alpha = 1$, then
(\ref{master equation}) reduces to the Gorini-Kossakowski-Sudarshan-Lindblad 
equation \cite{GoriniJMP76,LindbladCMP76}
and the equation of motion generates a completely positive trace-preserving (CPTP) 
channel
 \cite{KrausAnnPhys71,ChoiLin.Alg.App.75}.
If one or more $ \zeta_\alpha $ are negative, the map is either positive but not completely positive \cite{PechukasPRL94,ShajiPLA05,CarteretPRA08,13120908,150305342}, or else it fails to preserve positivity altogether (this is determined by a competition between terms with positive and negative $ \zeta_\alpha $).
In this paper we are concerned with both positive trace-preserving (PTP) channels \cite{SudarshanPR61} 
and CPTP channels. 
Substituting (\ref{master equation}) with dimension $d=2$ 
 into (\ref{bloch sphere picture}) yields
$ \frac{dr^\mu}{dt}  = {\rm tr} ( \sigma^\mu  
   [-i H , \rho ]  + \sum_\alpha
\zeta_\alpha \, \sigma^\mu  B_\alpha \rho B_\alpha^\dagger
+ \sigma^\mu  \{ L_{+} , \rho \} ) $.
Using 
$ \rho = \frac{ I_2 + r^\nu \sigma^\nu }{2} $
then leads to
\begin{eqnarray}
\frac{dr^\mu}{dt} = G^{\mu \nu}  r^\nu + 
C^\mu,  
\label{general pauli basis equation of motion}
\end{eqnarray}
where
\begin{eqnarray}
&&G^{\mu \nu} = - \epsilon^{\mu \nu \lambda} \, {\rm tr} ( \! H \sigma^\lambda  )
+ \sum_\alpha \frac{\zeta_\alpha}{2}  \,
 {\rm tr} ( \! \sigma^\mu B_\alpha \sigma^\nu B_\alpha^\dagger) 
+  {\rm tr} ( L_{+} \! ) \, \delta^{\mu \nu} \label{general pauli basis g}
\end{eqnarray}
and
\begin{eqnarray}
C^{\mu} = \sum_\alpha \! \frac{\zeta_\alpha }{2} \, {\rm tr} (  \sigma^\mu [ B_\alpha ,B_\alpha^\dagger] ).
\label{general pauli basis c}
\end{eqnarray}

\subsection{Jump operators}

Next we define our dissipation model by specifying a set $ \{ B_\alpha \}_{\alpha = 1}^7 $ 
of single-qubit jump operators.  
The dissipative torsion model in \cite{211105977} contains two  independently controlled nonunitary processes. The first is depolarization, requiring 
three jump operators. The second is a non-CP (meaning PTP but not CPTP) process detailed below, requiring four more. Therefore we need seven jump operators in total, which are given in Table \ref{jump operator table}. The signs 
$ \zeta_\alpha $ are also given. The jump operators used here are Hermitian so 
$C^\mu$ vanishes 
and the channel is unital (${\bf r} \! = \!  0$ is a fixed point).
Using the jump operators defined in Table \ref{jump operator table}, we obtain
\begin{eqnarray}
 L_{+} = - \frac{1}{2} \sum_\alpha \zeta_\alpha  B_\alpha^\dagger B_\alpha 
= - \frac{3\gamma}{8}  I_2 , \ \ {\rm tr}(L_{+}) = - \frac{3\gamma}{4}  .
\label{lplus}
\end{eqnarray}

\begin{table}[htb]
\centering
\begin{tabular}{|c|c|c|c|c|c|c|}
 \hline
$\alpha$ & $\zeta_\alpha$ & $B_\alpha$ & $B_\alpha^\dagger \! B_\alpha $ & $ \frac{1}{2} {\rm tr} ( \! \sigma^\mu B_\alpha \sigma^\nu B_\alpha^\dagger) $ \\
 \hline
1 &  +1 &  $\frac{\sqrt{\gamma}}{2} \sigma^x $ & $ \frac{\gamma}{4} I_2 $  & $ \gamma \! \begin{pmatrix} \frac{1}{4} & 0 & 0 \\ 0 & -\frac{1}{4} & 0 \\ 0 & 0 & -\frac{1}{4}  \end{pmatrix}$  \\
\hline
2 &  +1 &  $\frac{\sqrt{\gamma}}{2} \sigma^y $ & $ \frac{\gamma}{4} I_2 $  & $ \gamma \!  \begin{pmatrix} -\frac{1}{4} & 0 & 0 \\ 0 & \frac{1}{4} & 0 \\ 0 & 0 & -\frac{1}{4}  \end{pmatrix}$  \\
\hline
3 &  +1 &  $\frac{\sqrt{\gamma}}{2} \sigma^z $ & $ \frac{\gamma}{4} I_2 $ & $ \gamma \!  \begin{pmatrix} -\frac{1}{4} & 0 & 0 \\ 0 & -\frac{1}{4} & 0 \\ 0 & 0 & \frac{1}{4} \end{pmatrix}$   \\
\hline
4 &  +1 &  $ \sqrt{\frac{m}{8}}  (\sigma^x + \sigma^y + \sigma^z)  $ & $ \frac{3m}{8} I_2 $ & $ m \! \begin{pmatrix} -\frac{1}{8} & \frac{1}{4} & \frac{1}{4} \\ \frac{1}{4} & -\frac{1}{8} & \frac{1}{4} \\ \frac{1}{4} & \frac{1}{4} & -\frac{1}{8}  \end{pmatrix} $   \\
\hline
5 &  +1 &  $ \sqrt{\frac{m}{8}}  (\sigma^x - \sigma^y + \sigma^z)  $ & $ \frac{3m}{8} I_2  $  & $ m \! \begin{pmatrix} -\frac{1}{8} & -\frac{1}{4} & \frac{1}{4} \\ -\frac{1}{4} & -\frac{1}{8} & -\frac{1}{4} \\ \frac{1}{4} & -\frac{1}{4} & -\frac{1}{8}  \end{pmatrix} $  \\
\hline
6 &  -1 &  $ \sqrt{\frac{m}{8}} (\sigma^x + \sigma^y - \sigma^z)   $ & $\frac{3m}{8} I_2 $  & $ m \! \begin{pmatrix} -\frac{1}{8} & \frac{1}{4} & -\frac{1}{4} \\ \frac{1}{4} & -\frac{1}{8} & -\frac{1}{4} \\ -\frac{1}{4} & -\frac{1}{4} & -\frac{1}{8}  \end{pmatrix}$  \\
\hline
7 &  -1 &  $ \sqrt{\frac{m}{8}} (\sigma^x - \sigma^y - \sigma^z)   $ & $ \frac{3m}{8} I_2 $  & $ m \! \begin{pmatrix} -\frac{1}{8} & -\frac{1}{4} & -\frac{1}{4} \\ -\frac{1}{4} & -\frac{1}{8} & \frac{1}{4} \\ -\frac{1}{4} & \frac{1}{4} & -\frac{1}{8}  \end{pmatrix} $  \\
\hline
 \end{tabular}
 \caption{Jump operators used in the dissipative torsion model. 
The dimensionless model parameters $\gamma, m$ are real and nonnegative. $I_2$ is the two-dimensional identity.}
\label{jump operator table}
\end{table}

Having specified the linear model, we now include 
nonlinearity by using the state-dependent torsion Hamiltonian (\ref{torsion hamiltonian}) in the first term of (\ref{general pauli basis g}). This leads to a nonlinear ($z$ dependent) Pauli basis generator 
\begin{eqnarray}
G = m \, \lambda_4 - \gamma I_3 + 2 g z E_z,
\label{dissipative torsion model g}
\end{eqnarray}
where
\begin{equation}
\lambda_4 = 
\begin{pmatrix}
0 & 0 & 1 \\
0 & 0 & 0 \\
1 & 0 & 0 \\
\end{pmatrix} \!  
\ \ {\rm and} \ \ 
E_z = 
\begin{pmatrix}
0 & -1 & 0 \\
1 & 0 & 0 \\
0 & 0 & 0 \\
\end{pmatrix} \! .
\end{equation}
Here $\lambda_4$ is an SU(3) generator, $I_3$ is the three-dimensional identity, and
$E_z$ is an SO(3) generator. Dimensionless parameters $\gamma, m$ are real and nonnegative. Instead of  
(\ref{torsion model x})-(\ref{torsion model z}), we now have
\begin{eqnarray}
&& \frac{dx}{dt} = m z - \gamma x - 2g y z,  
\label{dissipative torsion model pauli basis x}  \\
&& \frac{dy}{dt} = - \gamma y + 2g x z, 
\label{dissipative torsion model pauli basis y}  \\
&& \frac{dz}{dt} = m x - \gamma z.
\label{dissipative torsion model pauli basis z} 
\end{eqnarray}

\clearpage

\subsection{Fixed points}

\begin{figure}
\begin{center}
\includegraphics[width=7.0cm]{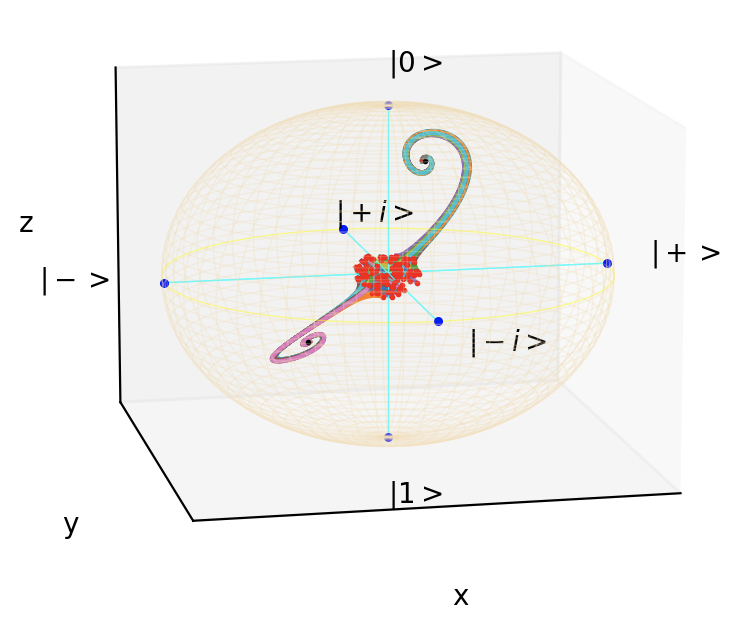} 
\caption{Curves show solutions of qubit equations of motion (\ref{dissipative torsion model pauli basis x})-(\ref{dissipative torsion model pauli basis z}) for 100 random initial conditions (uniformly distributed within a ball of radius 0.15), shown as red points. Blue points mark pure states on the  $x,y,z$ axes. Black dots are the fixed points
(\ref{stable fixed points}). Model parameters are $ \gamma = 0.5, \ m=1, \ {\rm and} \ g = 0.8$. In addition, $ \delta = 0.75 $ and $g_{\rm min} = 0.612$.} 
\label{attractors figure}
\end{center}
\end{figure} 

The fixed-point equations for the dissipative torsion model are
\begin{eqnarray}
m z - \gamma x - 2g y z = 0, \ \ 
 - \gamma y + 2g x z = 0, \ \ 
 m x - \gamma z = 0.
 \label{fixed point equations} 
\end{eqnarray}
The origin 
$ {\bf r}^{\rm fp}_{0} = (0,0,0) $ 
is a fixed point (the channel is unital), although not always stable. 
If $g=0$, all fixed points must be confined to the $y=0$ plane. Beyond  ${\bf r}^{\rm fp}_{0}$, there are no additional fixed points unless $\gamma =  m$, in which case there is a continuum of fixed points on the line ${\bf r}_{z=x} = \{ (w,0, w) : w \in {\mathbb R} \}$. The setting $\gamma = m$ is a singularity in the parameter space of the linearized model. Controlling  this singularity in the presence of nonlinearity is the key to autonomous QSD.

Next we assume $g \neq 0$ and obtain the fixed points $ {\bf r}^{\rm fp}_{\pm} $ and basins of attraction 
$ {\mathbb V}_{\pm} $. 
We have seen that 
$ {\bf r}^{\rm fp}_{0} = (0,0,0) $ 
is a fixed point. Are there more? Assuming $g > 0, \ m > 0, \ \gamma > 0, \ {\bf r} \neq 0$, and eliminating $z$, the fixed point conditions are
\begin{eqnarray}
m^2 - \gamma^2  = 2g m y, 
\label{first fixed point equation} \\
\gamma^2 y = 2gmx^2. 
\label{second fixed point equation}
\end{eqnarray}
Thus,  any additional fixed points must be confined to the plane
\begin{eqnarray}
y = \frac{m^2 - \gamma^2}{2gm} .
\end{eqnarray}
However (\ref{second fixed point equation}) requires $y \ge 0$, which is only possible when $m^2 \ge \gamma^2$.
Therefore, when $m^2  <  \gamma^2$, the only fixed point is ${\bf r}^{\rm fp}_{0}$ (and this is stable for all $m^2  <  \gamma^2$). But if
$ m > \gamma$,
the fixed point ${\bf r}^{\rm fp}_{0}$ is unstable and there is a new pair of stable fixed points at
\begin{eqnarray}
{\bf r}^{\rm fp}_{\pm} = \bigg( \! \pm \! \frac{ \gamma }{2g}
\sqrt{ \delta } , \ 
\frac{m}{2g} \, \delta , \
 \pm  \frac{m}{2g} 
\sqrt{ \delta } \bigg), \ \ 
 \delta := \frac{m^2 - \gamma^2}{m^2} \ge 0.
\label{stable fixed points}
\end{eqnarray}
For these to be contained within the Bloch sphere requires $g \ge  g_{\rm min}$,  where 
\begin{eqnarray}
g_{\rm min} = \sqrt{\frac{m^2 - \gamma^2}{2} } .
\label{gmin def}
\end{eqnarray}

\subsection{Autonomous state discrimination}

\begin{figure}
\begin{center}
\includegraphics[width=5.0cm]{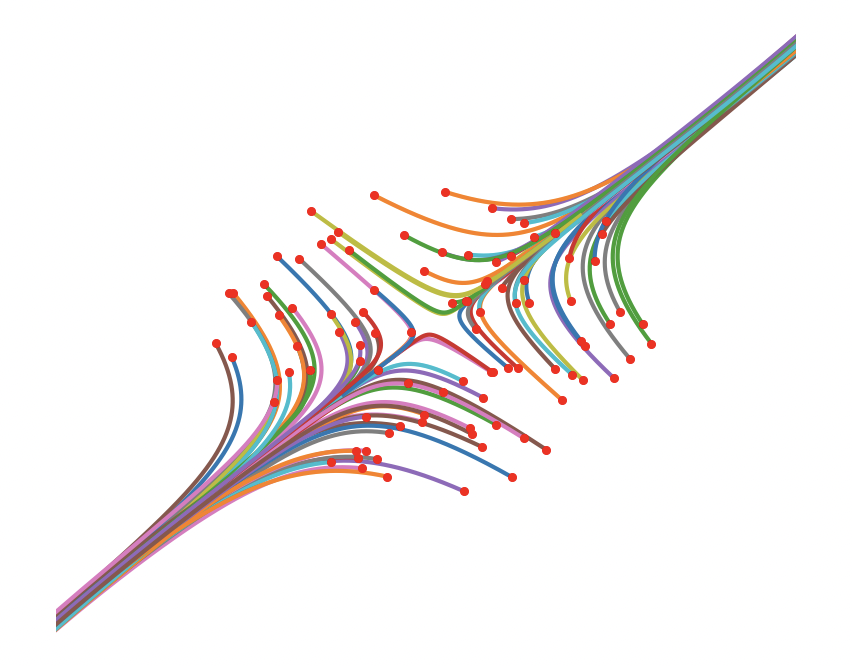} 
\caption{Expanded view of Figure~\ref{attractors figure}, seen from the $y$ direction, 
exposing the boundary separatrix between $ {\mathbb V}_{-} $ and $ {\mathbb V}_{+} $.
 Model parameters are $ \gamma = 0.5, \ m=1, \ {\rm and} \ g = 0.8$.} 
\label{separatrix figure}
\end{center}
\end{figure}

The autonomous QSD gate operates by placing 
the discriminator inputs 
$ \{ {\bf r}_{\rm a} , {\bf r}_{\rm b} \} $ 
exponentially close to the origin but flowing to different fixed points.
Figure \ref{attractors figure} shows the Bloch sphere together with the simulated evolution of random initial states (within a small radius and indicated by red dots) flowing to the discriminator outputs ${\bf r}^{\rm fp}_{\pm}$ (black dots).
To obtain the basins of attraction 
$ {\mathbb V}_{\pm} $, examine the equations close to the origin, where $| {\bf r} | \ll 1/g$ and the nonlinearity can be neglected. In this limit  
 the $y$ motion is decoupled from $x$ and $z$ [see (\ref{dissipative torsion model pauli basis y})],  and it is always stable for $\gamma > 0$. The $x$ and $z$ motion is simplest in rotated coordinates $ \xi_{\pm} = (z \pm x)/2$, where
\begin{eqnarray}
\frac{d\xi_{+}}{dt} = (m -\gamma) \xi_{+}
\ \ {\rm and} \ \ 
\frac{d\xi_{-}}{dt} = -  (m + \gamma) \xi_{-}.  
\label{xi pm equation}
\end{eqnarray}
The $\xi_{+}$ and $\xi_{-}$ variables are also decoupled in the linearized model. 
$\xi_{+}$ is the coordinate along the line $z=x$ (and $y=0$), and $\xi_{-}$ is the coordinate along the perpendicular line $z=-x$. Motion in the $\xi_{-}$ direction is always stable.
However the $\xi_{+}$ motion is unstable here because $m > \gamma$. 
Let \begin{eqnarray}
{\mathbb E}^2  = \{ (x,y,z) \in {\mathbb R}^3 : x + z = 0 \}  
\label{e plane}
\end{eqnarray}
 be the two-dimensional 
plane separating half-spaces $\xi_{+} < 0$ and $\xi_{+} > 0$. 
When $g=0$, the plane
${\mathbb E}^2$ separates $ {\mathbb V}_{-} $ (where $\xi_{+} < 0$)  and $ {\mathbb V}_{+} $ (where $\xi_{+} > 0$): it is the
separatrix in the linearized model.
In the nonlinear model, the separatrix is a non-planar surface, but in the region of interest near the origin the separatrix coincides with ${\mathbb E}^2$. Therefore ${\mathbb E}^2$ can be used as a surface across which the discriminator inputs 
$ \{ {\bf r}_{\rm a} , {\bf r}_{\rm b} \} $ are prepared.
Figure \ref{separatrix figure} shows  the evolution of states initialized close to the  plane $ {\mathbb E}^2$, separating the basins of attraction 
$ {\mathbb V}_{-} $ 
and $ {\mathbb V}_{+} $.

The autonomous QSD gate discussed here is no longer a single-input protocol because  the outputs $ {\bf r}^{\rm fp}_{\pm} $ are not perfectly distinguishable. However, because the fixed points $ {\bf r}^{\rm fp}_{\pm} $ are known explicitly [see (\ref{stable fixed points})], we expect that a second step mapping them to the perfectly distinguishable classical states $ \{ \KET{0}, \KET{1} \}$ can be implemented, making the gate both single-input and autonomous (we do not discuss the second step here).

\clearpage

\section{Experimental realization}
\label{experimental realization section}

It's not known whether the nonlinear qubit proposed  here is currently realizable with a two-component condensate,
but an optical cavity implementation has recently been demonstrated \cite{240219429}.
 One challenge is that condensates are subject to a variety of instabilities that result in new phases that may break assumptions of our model. For example, condensates with attractive interactions are subject to collapse to a liquid droplet phase \cite{PetrovPRL2015,170807806,SemeghiniFeriolPRL2018}, where the dilute and weakly interacting assumptions might be violated.
Two-component condensates with 
repulsive inter-component interaction  
$(a_{01} \! > \! 0) $
are also susceptible to phase separation \cite{HoShenoyPRL1996,AoChuPRB1998,LeeJorgensenPRA2016}, violating the assumption  (\ref{gaussian translational modes}) that the two components overlap.

Another challenge is that we want to control the torsion strength $g$ by varying an applied magnetic field $B$. However $g$ is determined by a combination $a_{00} + a_{11} - 2 a_{01}$ of scattering lengths,  each with its own resonance structure. Therefore, some luck is required to obtain a well-behaved $g(B)$. In this section we give a possible  implementation of the torsion model (\ref{torsion hamiltonian}) using condensed 
$^{39}$K atoms, identifying a stable region near $58 \, {\rm Gauss}$ where $g$ can be controlled and tuned close to zero.

\subsection{Choice of atom}

Choosing a bosonic qubit means choosing (i) a bosonic atom and (ii) a pair of robust internal states $\KET{\Psi_0} $ and $\KET{\Psi_1}$. 
$^{39}$K  has a nuclear spin of $I = \frac{3}{2}$ that couples with the electronic angular momentum $J=\frac{1}{2}$ to give a total  angular momentum ${\bf F} = {\bf I} + {\bf J}$. The ground state is split by the hyperfine interaction into an upper manifold with $F=2$ and a lower energy manifold with $F=1$. Here we consider a nonlinear qubit based on the $^{39}$K states \cite{170807806,SemeghiniFeriolPRL2018}
\begin{eqnarray}
\KET{\Psi_0} = \KET{ F\!=\!1, m_{F} \!=\! -1} , \ \ 
\KET{\Psi_1} = \KET{ F\!=\!1, m_{F} \!=\! 0} .
\label{qubit states}
\end{eqnarray}
Efficient all-optical trapping and cooling of $^{39}$K has been demonstrated \cite{SalomonFouchePRA2014,HerbstAlbersPRA2022}, leaving the magnetic field free to control the  scattering lengths. Stable miscible (homogeneous) phases of these components have been 
identified 
\cite{170807806,SemeghiniFeriolPRL2018},
and the states (\ref{qubit states})
lead to a well-behaved torsion strength $g(B)$ near $58 \, {\rm G}$, as detailed below. 

\begin{figure}
\begin{center}
\includegraphics[width=10.0cm]{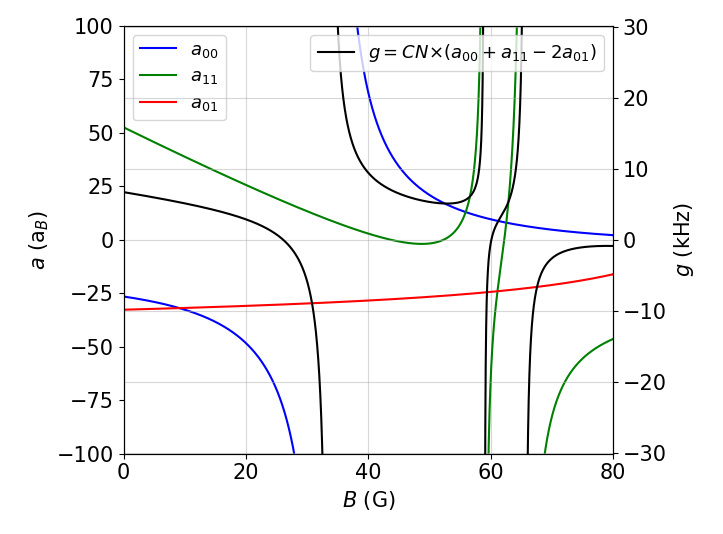} 
\caption{Scattering lengths versus magnetic field
for the $^{39}$K hyperfine states (\ref{qubit states}),  based on the Feshbach resonance model (\ref{feshbach model}), with parameters  calculated by Etrych {\it et al.} \cite{EtrychMartirosyanPRR23}. The blue, green, and red curves give the three scattering lengths  (left axis) in units of Bohr radius, $a_{\rm B}$.  The black curve gives the torsion strength (right axis) versus $B$, assuming $ N \! = \! 10^4$ atoms in a 3d harmonic trap with frequency $\nu_{\rm osc} 
\! = \! 500 \,  {\rm Hz}$. The coefficient $C$ is given in (\ref{c coefficient definition}). } 
\label{scattering length figure}
\end{center}
\end{figure}

Atoms in the electronic $\KET{\Psi_0}$ state
and $\KET{\Psi_1}$ state 
typically have different atomic interaction potentials and resulting collision behavior.
In the regime of interest here,
the atomic interactions are determined 
by s-wave scattering lengths $a_{00}$ and $a_{11}$ for elastic $\KET{\Psi_0}_A  + \KET{\Psi_0}_B $  and $\KET{\Psi_1}_A + \KET{\Psi_1}_B$ collisions, as well as by the s-wave scattering length $a_{01}$  for elastic $\KET{\Psi_0}_A + \KET{\Psi_1}_B $  collisions (subscripts $A$ and $B$ label the colliding atoms). 
The scattering lengths appear in the  Hamiltonian (\ref{two boson hamiltonian}) via the interaction parameters (\ref{U definitions}), and they are tunable near Feshbach resonances, as shown in Figure \ref{scattering length figure}.

The Feshbach resonance model for an s-wave scattering length is characterized by a triple $(B_{\rm res} , \Delta ,a_{\rm bg} )$ specifying the center $B_{\rm res}$, width $\Delta$, and background value $a_{\rm bg}$ for each resonance \cite{08121496}. A resonance is assumed  to modify the background scattering length by a factor
$ 1 - \frac{\Delta}{B - B_{{\rm res}} }  $, 
which has  a simple pole at $B_{\rm res}$.
For non-overlapping resonances (assumed here) they combine as
\begin{eqnarray}
a(B) = a_{\rm bg}(B) \prod_j  \bigg( 1 - 
\frac{\Delta_j}{B - B_{{\rm res},j} } \bigg),
\label{feshbach model}
\end{eqnarray}
where $a_{\rm bg}(B)$ is a smooth background obtained by interpolating the $a_{\rm bg}$ values. We use $^{39}$K Feshbach resonance parameters calculated by Etrych {\it et al.} \cite{EtrychMartirosyanPRR23}
and summarized below in Table \ref{feshbach model table}. The resulting scattering lengths (blue, green, and red curves) are plotted in Figure \ref{scattering length figure}. 

The scattering lengths 
plotted in Figure \ref{scattering length figure}
feature a broad resonance in $a_{00}$ at approximately $ 34 \, {\rm G}$, and a pair of narrow resonances in $a_{11}$ near
$ 59 \, {\rm G}$ and $ 66  \, {\rm G}$.
An $a_{01}$ resonance at $ 114  \, {\rm G}$ 
(out of the range of the plot)  produces a slowly varying attractive $a_{01}$, 
stabilizing the homogeneous phase.

Next, we sketch an approximate density profile of the $^{39}$K condensate using the Thomas-Fermi approach,  adapted  to a two-component condensate \cite{HoShenoyPRL1996,AoChuPRB1998}.
Applying the mean field approximation 
$ \phi_\alpha({\bf r}) \rightarrow \Psi_\alpha({\bf r})
:= \langle \phi_\alpha({\bf r}) \rangle $
directly to (\ref{two boson hamiltonian}),
with $\Omega = 0$, leads to
\begin{eqnarray}
&& E[\Psi_\alpha]  =  \int \! \! d^3r   
  \bigg\lbrace  \! \sum_{\alpha=0,1} \bigg[ 
 \frac{ | \nabla \Psi_\alpha |^2 }{2m} \! + \!
 V_\alpha({\bf r}) \,  |\Psi_\alpha({\bf r})|^2 +
 \frac{U_{\alpha \alpha}}{2}  |\Psi_\alpha({\bf r})|^4
 \bigg]  \, + \  U_{01}  |\Psi_0({\bf r})|^2  |\Psi_1({\bf r})|^2  \bigg\rbrace  \nonumber \\
&& \ \ \ \ \ \ \ \ \ - \mu_0 N_0  - \mu_1 N_1 ,
\label{two boson energy functional}
 \end{eqnarray}
where chemical potential terms are included  to control $N_0$ and $N_1$. 
The order parameters 
$\Psi_{\alpha}({\bf r}) \, (\alpha \in \{0,1\} )$
are normalized according to
$\int \! d^3r \,  |\Psi_\alpha({\bf r})|^2 = N_\alpha $.
In the Thomas-Fermi approach, the kinetic energy term in  (\ref{two boson energy functional}) is neglected, and the energy is rewritten as a functional of the real-space atom number densities 
$ n_{0}({\bf r}) =   |\Psi_0({\bf r})|^2 $
and
$n_{1}({\bf r}) =   |\Psi_1({\bf r})|^2 $, as
 \begin{eqnarray}
 E[n_\alpha]  =  \int \! \! d^3r   
 \bigg\lbrace \!  \sum_{\alpha=0,1} \bigg[ \bigg(
 V_\alpha({\bf r}) \!  - \! \mu_\alpha \bigg) 
n_{\alpha}({\bf r}) +
 \frac{U_{\alpha \alpha}}{2} [n_{\alpha}({\bf r})]^2 
 \bigg]   +   U_{01}  n_{0}({\bf r}) n_{1}({\bf r})  \bigg\rbrace .
\label{two boson density functional}
 \end{eqnarray}
 Extremizing the energy leads to the coupled equations
\begin{eqnarray}
\frac{ \delta  E[ n_{\alpha} ] }{ \delta n_{0}({\bf r})  } = \bigg(
 V_0({\bf r}) \!  - \! \mu_0 \bigg) 
+  U_{00}  \, n_{0}({\bf r}) 
+   U_{01}  \,  n_{1}({\bf r}) = 0 , \nonumber \\
\frac{ \delta  E[ n_{\alpha} ] }{ \delta n_{1}({\bf r})  } = \bigg(
 V_1({\bf r}) \!  - \! \mu_1 \bigg) 
+  U_{11}  \, n_{1}({\bf r}) 
+   U_{01}  \,  n_{0}({\bf r}) = 0.
\label{coupled tf equations}
\end{eqnarray}

\clearpage

\begin{figure}
\begin{center}
\includegraphics[width=9.0cm]{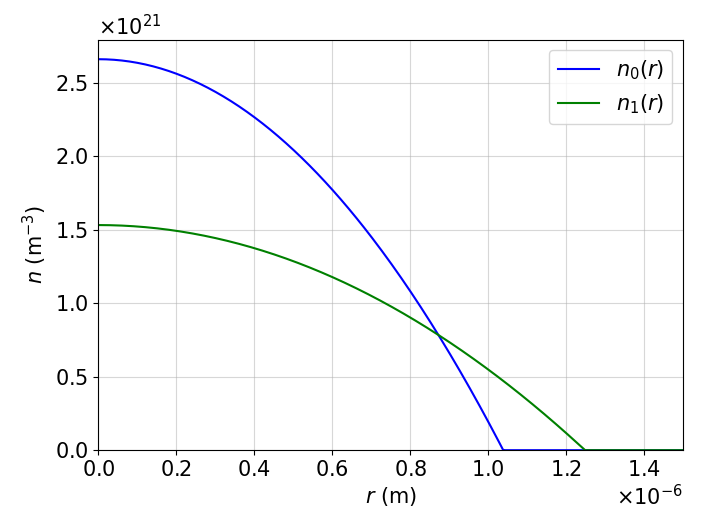} 
\caption{Thomas-Fermi density profiles for a two-component $^{39}$K condensate at $B \! = \! 58 \, {\rm G}$. Here $n_0$ and $n_1$ are the densities of atoms in the states (\ref{qubit states}), versus radial distance from the center of the 3d harmonic trap with $\nu_{\rm osc} \! = \! 500 \, {\rm Hz}$.} 
\label{density profile figure}
\end{center}
\end{figure}

A solution to the coupled  equations (\ref{coupled tf equations}) is
\begin{eqnarray}
 n_{0}({\bf r})  \! = \! \frac{  U_{11}   [ \mu_0 \! - \!  V_0({\bf r})]   -  U_{01} \,  [ \mu_1 \! - \!  V_1({\bf r})]  }{U_{00} U_{11} - U_{01}^2 } 
 \  \  {\rm and} \  \
 n_{1}({\bf r}) \!  = \! \frac{  U_{00} \, [ \mu_1 \! - \!  V_1({\bf r})]  -  U_{01} \,  [ \mu_0 \! - \!  V_0({\bf r})]  }{U_{00} U_{11} - U_{01}^2 } .  \ \ \ 
 \label{density profile equations}
 \end{eqnarray}
 To evaluate these expressions, 
 it is convenient to write the interaction parameters  (\ref{U definitions}) as
 $ U_{\alpha \beta} =  ( \frac{ a_{\alpha \beta} }{ a_{\rm B} } ) U_{\rm B} $,
 with
$ U_{\rm B} \! = \!  \frac{ 4 \pi \hbar^2  a_{\rm B}}{m} \! = \!  1.143  \times 10^{-52} 
\  {\rm J \, m^3 } $ 
(using $m \! = \!  6.47 \! \times \!  10^{-26} \, {\rm kg}$ for $^{39}$K). 
In the example considered here, there are $N = 10^{4}$ atoms total ($N_0 \!=\! N_1 \!= \! 5000$), and both trapping potentials $V_{\alpha} = \frac{1}{2} m \omega_{\alpha}^2 r^2 $  have the same frequency $\omega = 2 \pi \nu_{\rm osc}$, with $ \nu_{\rm osc} \! = \!  500 \, {\rm Hz}$.
The density profiles 
(\ref{density profile equations}) are plotted in Figure~\ref{density profile figure}.

\vspace{0.5in}

\begin{table}[htb]
\centering
\begin{tabular}{|c|c|c|c|c|}
 \hline
Scattering length & Collision & $B_{\rm res} \, ({\rm G})$ & $\Delta \, ({\rm G}) $ & $a_{\rm bg} \, (a_{\rm B}) $  \\
\hline
$a_{00}$ & $ \KET{1,-1}_A + \KET{1,-1}_B $ 
& 33.568 & 79.469 & -13.50   \\
 &  & 162.347 & -60.628 & -11.73   \\
  &  & 560.935 &  -55.358 & -29.10  \\
 \hline
 $a_{11}$ & $ \KET{1,0}_A + \KET{1,0}_B $ 
 & 58.949 &  -6.648 & -29.22  \\
 &  & 65.573 & -3.361 &  -41.87  \\
 &  & 472.118 & -117.78 & -16.96  \\
  &  & 490.930 & -1.045 & -133.43  \\
  \hline
 $a_{01}$ & $ \KET{1,-1}_A + \KET{1,0}_B $ 
 & 113.768 &  -19.215 & -39.33  \\
 &  & 525.995 & -28.249 & -31.04  \\
 \hline
\end{tabular}
\caption{Feshbach resonance parameters
for $^{39}$K. Here $ \KET{F,m_F}_A + \KET{F^\prime,m_F^\prime}_B$
 denotes a collision between atom A in hyperfine state $\KET{F,m_F}$ and atom B in state 
$ \KET{F^\prime,m_F^\prime}$.
Model parameters are obtained from Table IV of 
Ref.~\cite{EtrychMartirosyanPRR23}.}
\label{feshbach model table}
\end{table}

\clearpage

\subsection{Nonlinear coupling $g(B)$}

The coupling $g$ defined in (\ref{b and g}) 
gives the strength of the torsion.
In particular, the nonlinear Hamiltonian
(\ref{torsion hamiltonian}) produces a
$z$ rotation of a Bloch vector ${\vec r} = (x,y,z)$
by a frequency $2gz$ that depends on its $z$ coordinate.  Using  (\ref{u definitions}) and (\ref{k definitions}) leads to
\begin{eqnarray}
g = \frac{N}{4} \bigg(
\frac{U_{00} }{  ( 2 \pi \ell^2_{0} )^\frac{3}{2}  }
+ \frac{U_{11} }{  ( 2 \pi \ell^2_{1} )^\frac{3}{2}  } 
- 2 \frac{U_{01}}{ [ \pi (\ell_0^2 \! + \! \ell_1^2) ]^\frac{3}{2}   }  \bigg) .
\label{g in terms of u}
\end{eqnarray}
Here $N$ is the total number of atoms, the $U_{\alpha \beta}$ are interaction parameters
(\ref{U definitions}) appearing in the many-body Hamiltonian (\ref{two boson hamiltonian}), and 
the $\ell_\alpha$ are harmonic oscillator length scales for the Gaussian spatial modes (\ref{gaussian translational modes}).
Expressing this in terms of scattering lengths leads to
\begin{eqnarray}
 g =  CN \bigg(  \frac{ a_{00} + l^3 \, a_{11} - 2 \big( \frac{2}{1 + l^{-2}} \big)^{\! \frac{3}{2}} \, a_{01}}{ a_B} \bigg) ,
  \ \ l = \frac{l_0}{ l_1},
\ \  C =  \pi \sqrt{ \hbar m} \, a_{\rm B} \, \nu_{{\rm osc}}^\frac{3}{2} .
\label{c coefficient definition}
\end{eqnarray}
Here $a_{\rm B}$ is the Bohr radius.
For a given atomic mass $m$, $C$ depends only on the trap frequency $\nu_{{\rm osc}} $
for the $\KET{\Psi_0}$-state atoms. 
Note that
$\ell_0 = \sqrt{ \hbar / m \omega_0 }$, where $\omega_0 = 2 \pi \nu_{\rm osc}$.
For a $\nu_{\rm osc} \! = \! 500 \, {\rm Hz}$ trap, we have
$\ell_0 = 0.72 \, \mu {\rm m} $ and
$C/h \! = \! 0.00733 \, {\rm Hz}$.
The nonlinear coupling for a symmetric trap ($\ell_0 = \ell_1)$ is also plotted in Figure \ref{scattering length figure}. 

Why does $g$ depend linearly on $N$? This is a consequence of the large $N$ limit in (\ref{k definitions}). We have seen that the nonlinear picture applies to a specific limiting case of the condensate, where $N \rightarrow \infty$ and  $u_{\alpha \beta} \rightarrow K_{\alpha \beta}/N$ with $K_{\alpha \beta}$ constant
(here $\alpha, \beta \in \{0,1\}$).
The quantities $K_{\alpha \beta}$ appearing in the nonlinear model are therefore proportional to an interaction parameter $u_{\alpha \beta}$  (or scattering length $a_{\alpha \beta}$), 
multiplied by $N$. To keep the product $N(a_{00} + a_{11} - 2   \, a_{01})$  constant as $N$ increases, we must tune $a_{00} + a_{11} - 2   \, a_{01}$
closer and closer to zero to remain in the 
nonlinear mean field regime.

\vspace{0.3in}

\begin{table}[htb]
\centering
\begin{tabular}{|c|c|}
 \hline
$a_{00}$  &  $11.06 \,  a_{\rm B}$  \\
$a_{11}$  &  $65.35 \,  a_{\rm B} $  \\
$a_{01}$  &  $ -24.91 \,  a_{\rm B}$  \\
 \hline
$ \sqrt{a_{00} a_{11}}   $  &  $26.88  \,  a_{\rm B} $  \\
$ (a_{00} \! + \! a_{11})/2  $  &  $38.20  \,  a_{\rm B} $  \\
$ a_{00} a_{11}  \! - \! a_{01}^2 $  &  $ 101.88 \, a_{\rm B}^2$  \\
$ a_{00} + a_{11} \! - \!   2 a_{01} $  &  $126.23  \,  a_{\rm B}$  \\
 \hline
\end{tabular}
\caption{$^{39}$K s-wave scattering lengths and their characteristics at $B \! = \! 58 \, {\rm G}$, according to the Feshbach resonance model (\ref{feshbach model})  with parameters from Etrych {\it et al.} \cite{EtrychMartirosyanPRR23}.
The subscripts $0,1$ refer to the hyperfine states (\ref{qubit states}). $a_{\rm B}$ is the Bohr radius.}
\label{scattering length table}
\end{table}

\begin{figure}
\begin{center}
\includegraphics[width=10.0cm]{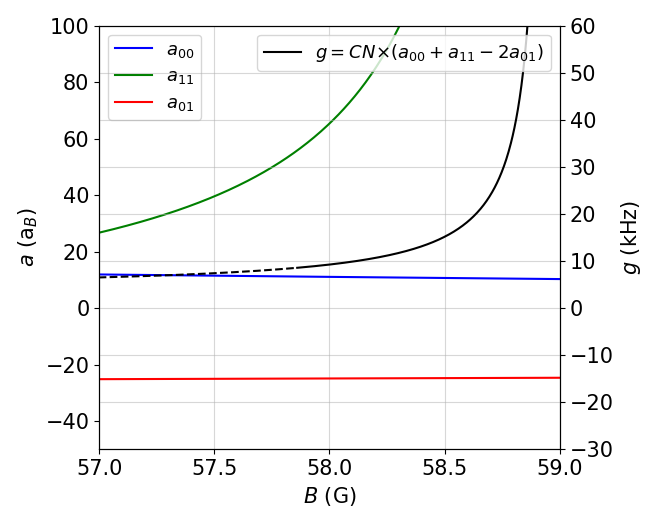} 
\caption{Expanded view of the coupling 
(right axis) near $58 \, {\rm Gauss}$. The dotted line indicates a region where the condensate collapses to a  droplet.  The scattering lengths at $B \! = \!  58 \, {\rm G}$  are given in Table \ref{scattering length table}.} 
\label{expanded figure}
\end{center}
\end{figure}

We turn now to a discussion of the condensate properties in the magnetic field region near $58 \, {\rm G}$.  According to the Feshbach resonance model, the scattering lengths at $B = 58 \, {\rm G}$  have the values listed in Table \ref{scattering length table}. An expanded view of the torsion strength near $58 \, {\rm G}$ is shown in Figure \ref{expanded figure}. In the region of magnetic field shown, the scattering lengths $a_{00}$ and $a_{11}$ are both positive, whereas $a_{01}$ is negative. In this situation, the condensate is stable as long as $a_{00}$ and  $a_{11}$ are large enough to counter the attractive $a_{01}$. 
A mean-field estimate for the 
stability condition is \cite{PetrovPRL2015}
\begin{eqnarray}
- \sqrt{a_{00} a_{11}}  < a_{01} .
\label{stability condition against droplet}
\end{eqnarray}

\clearpage

According to the scattering lengths given in Table \ref{scattering length table}, the condensate is predicted to be stable at $58 \, {\rm G}$.
If $a_{01}$ becomes sufficiently negative as to violate the condition (\ref{stability condition against droplet}), however, the gas collapses into a liquid droplet, stabilized against further collapse by  kinetic and correlation energy \cite{PetrovPRL2015,170807806,SemeghiniFeriolPRL2018}. In Figure \ref{expanded figure}, the nonlinear coupling in the unstable region is shown as a dotted line. We see that as the field is decreased, the condensate becomes unstable to droplet formation around 
\begin{eqnarray}
B_{\rm c} =   57.9 \, {\rm G} \  ({\rm model}),
\label{bc value theory}
\end{eqnarray}
mainly due to the sharp decrease in $a_{11}$.
Above $B_{\rm c}$ the coupling varies smoothly.

In conclusion,  it appears that a $^{39}$K condensate based on the states (\ref{qubit states}) can be used to implement a nonlinear qubit with a tunable (magnetic-field dependent)  torsion strength near $58 \, {\rm G}$.
It should be emphasized, however, that there are significant uncertainties in the Feshbach resonance model used here, both in the functional form (\ref{feshbach model}) and in the model parameters 
$ \{ B_{{\rm res},j} , \Delta_j ,a_{{\rm bg},j}  \}$
listed in Table \ref{feshbach model table}.
For example, there are differences (in some cases significant) between the $^{39}$K parameters used in 
Refs.~\cite{PetrovPRL2015,170807806,SemeghiniFeriolPRL2018}. 
These also differ from parameters calculated  by  Etrych {\it et al.} \cite{EtrychMartirosyanPRR23}, 
which are based on 
new $^{39}$K potentials of Tiemann {\it et al.} 
\cite{TiemannGersemaPRR2020}
and the spin-orbit coupling function of
Xie {\it et al.} \cite{XieVandeGraaffPRL2020}.
One way to gauge the accuracy of the scattering lengths near $58 \, {\rm G}$ is to compare the critical field (\ref{bc value theory}),
below which the condensate collapses,
to the measured value 
\cite{170807806,SemeghiniFeriolPRL2018}:
\begin{eqnarray}
B_{\rm c} =  56.9 \, {\rm G} \ ({\rm experiment}).
\label{bc value experiment}
\end{eqnarray}
In the future it would also be useful to assess the accuracy of the coupling $g(B)$.

\clearpage

\section{Discussion}
\label{conclusions section}

In standard gate-based quantum computation, 
a register of qubits is initialized to a product 
state, such as $\KET{0}^{ \! \otimes \NUMBER} \! ,\, $ after which  gates (linear CPTP channels) are applied, entangling the qubits.
In the nonlinear approach discussed here, a register of bosonic qubits is initialized into a symmetric product state through condensation and subsequently controlled by varying the qubit-qubit interaction. Entanglement is suppressed by making $\NUMBER$ large and the interaction weak, so
 the qubits ideally remain in a product state  $\KET{\psi(t)}^{ \! \otimes \NUMBER} $ throughout the computation. The $\NUMBER$ atoms  simulate one nonlinear qubit.

We considered a setup for a two-component BEC to implement  a nonlinear qubit with torsion (one-axis twisting). Torsion is computationally powerful because it violates trace-distance monotonicity and allows a pair of states to become more distinguishable.
Our work builds on previous proposals \cite{240416288,240310102}  for creating nonlinear qubits, but is likely simpler as it is based on previously established spin-squeezing techniques \cite{\SPINSQUEEZING}.

The nonlinear picture is intriguing, but it is also important to keep in mind its limitations: 
(i) The reduction of 3SAT to QSD requires that  the nonlinear qubit be entangled with a scalable, error-corrected quantum computing architecture, such as superconducting qubits \cite{211203708,220706431,231105933}, neutral atom arrays \cite{BluvsteinNAT22,GrahamNat22}, or trapped ions  \cite{EganDebroyNat2002,230503828}. A promising approach is to couple the BEC to a trapped ion \cite{GerritsmaNegrettiPRL2012,JogerNegrettiPRA2014,EbghaSaeidianPRA2019}, 
building on recent 
advances combining cold atom magneto-optical traps and optical lattices with electromagnetic traps for ions 
\cite{IdziaszekCalarcoPRA2007,TomzaJachymskiRMP2019,JyothiEgodapitiyaRSI2019,KarpaAtoms2021,LousGerritsmaAdvances2022,ZipkesPalzerNat2010,SchmidtWeckesserPRL2020,GerritsmaNegrettiPRL2012,JogerNegrettiPRA2014,EbghaSaeidianPRA2019}. 
(ii) The ability to implement the Viviani QSD gate does not imply
the ability to solve 3SAT because that requires a different QSD gate
(\ref{discrimination problem for 3sat}), which 
maps a large set of potential inputs to $\KET{1}$.
(iii) The nonlinear approach trades exponential time complexity for space complexity, requiring $\NUMBER$ (the number of atoms) to be large. Long computations will require $\NUMBER$ to be exponentially large, eventually limiting scalability.  
(iv) Long computations will also require error correction. This is an interesting and challenging problem, as existing quantum error correction techniques assume linear quantum mechanics.

\acknowledgements

This work was partly supported by the NSF under grant no.~DGE-2152159. It is a pleasure to thank Yohannes Abate and Victoria Ordonez for useful discussions.

\appendix

\section{Entanglement monogamy}
\label{entanglement monogamy section}

In this section we give an elementary derivation of the monogamy inequality (\ref{monogamy}) 
for three qubits in a pure state
 $\KET{\psi}$,
\begin{eqnarray}
 \tau( \rho_{12} )  +  \tau( \rho_{13} ) 
  \le 4 \, {\rm det}(\rho_1)  , \ \  
  \rho_1 = {\rm tr}_{23}(\rho_{123} ) , \ \ 
  \rho_{123} = \KETBRA{\psi}, 
 \label{n=3 monogamy}
\end{eqnarray}
originally proved in \cite{CoffmanPRA00}, but now simplified by the additional assumption of permutation symmetry. 
Here ${\rm tr}_{23}(\cdot)$ denotes partial trace over qubits 2 and 3. Concurrence \cite{WootersPRL98} and tangle (concurrence squared) measure the entanglement in some two-qubit pure or mixed state $\rho_{ij} \in {\mathbb C}^{4 \times 4}$ through a matrix 
\begin{eqnarray}
T_{ij} := \rho_{ij} \, {\tilde \rho_{ij}} , \ \  
{\tilde \rho_{ij}} := 
\sigma^y \! \otimes \! \sigma^y \,  {\bar \rho}_{ij} \,\sigma^y \! \otimes \! \sigma^y , \ \ 
\text{(for qubits} \ i \text{ and } j)
 \label{t matrix}
\end{eqnarray}
where ${\bar \rho}_{ij} $ is the complex conjugate of $\rho_{ij}$ in the standard ($\sigma^z$ eigenstate) basis. 
Typically the matrix $T_{ij}$  depends on the qubit pair $(i,j)$ considered. 
Concurrence $C(\rho_{ij}) $ is defined through the eigenvalues of $T_{ij}$, which are nonnegative with square roots $\lambda_i$:
\begin{eqnarray}
{\rm spec}(T_{ij}) = \{ \lambda^2_1, \lambda^2_2, \lambda^2_3,   \lambda^2_4 \} , \ \ 
 \lambda_1 \ge  \lambda_2 \ge  \lambda_3 \ge  \lambda_4 \ge 0.
 \label{t spectrum}
\end{eqnarray}
Then \cite{WootersPRL98}
\begin{eqnarray}
\tau(\rho_{ij}) = C(\rho_{ij})^2, \ \ 
C(\rho_{ij}) = \max \{  \Delta , 0 \} ,
\ \  \Delta := \lambda_1 - \lambda_2 -  \lambda_3 -  \lambda_4 .
 \label{tangle definiition}
\end{eqnarray}
Entangled and separable states are divided by a hyperplane $\Delta = 0$ in the space of allowed $\lambda_i$. Entangled states live in the open half-space $\Delta > 0$, while separable states (irrelevant to monogamy) live in the closed half-space $\Delta \le 0$. However our proof of (\ref{n=3 monogamy}) 
will require $\Delta \ge 0$, so that $C(\rho_{ij}) = \Delta $. We achieve this requirement by assuming that $\Delta > 0 $ until the end of the calculation, when the limit  $\Delta \rightarrow 0 $ can be taken. With this assumption,
\begin{eqnarray}
\tau(\rho_{ij}) =  (\lambda_1 - \lambda_2 -  \lambda_3 -  \lambda_4)^2 \le 
\lambda^2_1+  \lambda^2_2 +  \lambda^2_3 +   \lambda^2_4 = {\rm tr}(T_{ij}) .
\label{t spectrum inequality}
\end{eqnarray}

To derive (\ref{n=3 monogamy}), first note that every pure symmetric state of three qubits lies in a 4-dimensional subspace, namely
\begin{eqnarray}
&& \KET{\psi_{\rm s} } = a \, \KET{000} + b \, \KET{111}
+ c \, \KET{W} + d \, \KET{V}, \ \ 
|a|^2 + |b|^2  + |c|^2  + |d|^2 = 1,
\end{eqnarray}
where 
\begin{eqnarray}
 \KET{W} = \frac{ \KET{001} + \KET{010} + \KET{100} }{\sqrt 3} \ \ {\rm and} \ \ 
\KET{V} = \frac{ \KET{011} + \KET{101} + \KET{110} }{\sqrt 3}
\end{eqnarray}
are both permutation symmetric.
Our derivation will also make use of the following two-qubit states
\begin{eqnarray}
&& \KET{A} =  a  \KET{00} + \frac{d}{\sqrt{3}} \, \KET{11} + \sqrt{ \frac{2}{3} } \, c \,\KET{\Psi^+} , \nonumber \\  
&& \KET{B} =  \frac{c}{\sqrt{3}} \,  \KET{00} + b \, \KET{11} + \sqrt{ \frac{2}{3} } \, d \,\KET{\Psi^+} ,  \nonumber \\
&& \KET{C} =  \frac{\bar d}{\sqrt 3} \,  \KET{00} 
+ {\bar a}  \,  \KET{11}  - \sqrt{ \frac{2}{3} }  \, {\bar c}  \, \KET{\Psi^+}, \nonumber \\
&& \KET{D} =  {\bar b}  \,  \KET{00} 
+ \frac{\bar c}{\sqrt 3}  \,  \KET{11}  - \sqrt{ \frac{2}{3} }  \, {\bar d}  \, \KET{\Psi^+},
 \end{eqnarray}
where $ \KET{\Psi^+} = 2^{-\frac{1}{2} } (\KET{01}  +  \KET{10}) $
is a Bell state and ${\bar z }$ denotes complex conjugation. A short calculation then shows that 
 \begin{eqnarray}
&  | \langle A | C \rangle |^2 
+  | \langle A | D \rangle |^2
+  | \langle B | C \rangle |^2
+  | \langle B | D \rangle |^2 & \nonumber \\
& = 2 |a|^2 |b|^2 + \frac{4}{3} ( |a|^2 |d|^2 + |b|^2 |c|^2 )
+ \frac{2}{9}  |c|^2 |d|^2   
+ \frac{4}{9} ( |c|^4  +|d |^4 ) 
- \frac{4}{3} \, {\rm Re} [a b {\bar c} {\bar d} 
+ \frac{2}{\sqrt 3}(  a {\bar c}^2 d + b c {\bar d}^2  ) ]  & \nonumber  \\
&  = 2 (  |a|^2 + \frac{2}{3} |c|^2 + \frac{1}{3} |d|^2  ) ( |b|^2 +  \frac{1}{3} |c|^2 + \frac{2}{3} |d|^2  )
- 2 |  \frac{1}{\sqrt 3} a {\bar c}   +  \frac{2}{3} c {\bar d} +  \frac{1}{\sqrt 3} {\bar b} d|^2 .&
\label{main monogamy identity}
\end{eqnarray}
To obtain (\ref{n=3 monogamy}) note that
\begin{eqnarray}
 \rho_{12}  = {\rm tr}_3(\KETBRA{\psi_{\rm s} } )  =  \KETBRA{A} \! + \! \KETBRA{B}  
 \ \ {\rm and} \ \ 
 {\tilde \rho}_{12}  = 
\KETBRA{C}  \! + \! \KETBRA{D} \! , 
\end{eqnarray}
where ${\rm tr}_{3}( \cdot ) $
is partial trace over qubit 3. Furthermore,
\begin{eqnarray}
 \rho_{1} =   {\rm tr}_2( \rho_{12} ) =
\begin{pmatrix} |a|^2 + \frac{2}{3} |c|^2 + \frac{1}{3} |d|^2 & \frac{1}{\sqrt 3} a {\bar c}   +  \frac{2}{3} c {\bar d} +  \frac{1}{\sqrt 3} {\bar b} d  \\ 
 \frac{1}{\sqrt 3} {\bar a} c +  \frac{2}{3} {\bar c} d +  \frac{1}{\sqrt 3} b  {\bar d}  & 
|b|^2 +\frac{1}{3} |c|^2 +  \frac{2}{3} |d|^2   
\end{pmatrix} \! .
\end{eqnarray}
Then
\begin{eqnarray}
{\rm tr}( \rho_{12}  {\tilde \rho}_{12} ) = | \langle A | C \rangle |^2 
+  | \langle A | D \rangle |^2
+  | \langle B | C \rangle |^2
+  | \langle B | D \rangle |^2 
= 2 \, {\rm det}(\rho_1) .
\end{eqnarray}
Using permutation symmetry we also have
$ {\rm tr}( \rho_{13}  {\tilde \rho}_{13} ) = 2 \, {\rm det}(\rho_1)$. Thus, inequality (\ref{t spectrum inequality}) leads to
$ \tau(\rho_{12})  + \tau(\rho_{13}) \le   {\rm tr}( \rho_{12}  {\tilde \rho}_{12} ) + {\rm tr}( \rho_{13}  {\tilde \rho}_{13} )  = 4 \, {\rm det}(\rho_1) $
as required.

Three qubits is the minimum number needed to analyze the sharing of two-qubit entanglement. 
Coffman, Kundu, and Wootters \cite{CoffmanPRA00} derived (\ref{n=3 monogamy}) and conjectured that it would generalize to (\ref{monogamy}) for $\NUMBER>3$. This was later proved by Osborne and Verstraete \cite{OsborneVerstraetePRL2006}, an important milestone in monogamy theory \cite{HorodeckiRMP09,220100366}.
We do not provide an independent proof for general $\NUMBER$, but rely on Osborne and Verstraete \cite{OsborneVerstraetePRL2006}
for this justification.

\section{Monotonicity of trace distance}
\label{trace distance monotinicity section}

Here we prove the monotonicity of trace distance 
given above in (\ref{trace distance monotinicity}),
which governs how the distance between a pair of states can change as they evolve under the same positive trace-preserving (PTP) channel $\phi$.
The trace norm $ \| \cdot  \|_1 $ is defined in (\ref{trace norm definition}). We will also use an equivalent variational characterization of the trace norm on Hermitian operators \cite{WildeQuantumInformation}
\begin{eqnarray}
\| X \|_1 =  2 \max_{ 0 \preceq E \preceq I}  {\rm tr}(EX) = 2 \, {\rm tr}(E_{\star}  X), 
\label{trace norm variational}
 \end{eqnarray}
where the maximization is over all positive semidefinite (PSD) operators $E$ with spectra bounded by unity (such as POVM elements and projectors). In the second expression,
$ E_{\star} :=  {\rm argmax}_{ 0 \preceq E \preceq I}  {\rm tr}(EX) $ is the optimal $E$. 
In addition, the norm of any state is unity: 
$ \| \rho \|_1 = {\rm tr}( |\rho | ) = {\rm tr}( \rho)
= 1$. The inequality (\ref{trace distance monotinicity})  means that linear PTP channels are either distance-preserving (unitary) or contractive, but never expansive. It expresses the monotonicity of trace distance under PTP maps.
Although (\ref{trace distance monotinicity}) is a special case of a theorem on the monotonicity of a (generalized) relative entropy \cite{RuskaiRMP1994}, we  repeat  a simpler direct proof \cite{WildeQuantumInformation,Neilsen2000} here.
Given a pair of states $\rho$ and $\rho'$, let $ X = \rho - \rho' $ be a Hermitian operator. Then
\begin{eqnarray}
&& X = \rho - \rho' = U D U^\dagger =  U  \big( D_{>} -  D_{<}  \big) U^\dagger  = X_{>} - X_{<} , \\
&& X_{>}  =  U  D_{>} U^\dagger , \ \ X_{<}  =  U  D_{<} U^\dagger , \ \ X_{>} X_{<} = X_{<} X_{>}  = 0 ,
\end{eqnarray}
where $U$ is unitary, $D$ is diagonal,  and where $D_{>} $ and $D_{<} $ are diagonal and PSD. Here $D_{>} $ contains the positive eigenvalues of $X$ and $D_{<} $ contains the absolute values of the negative eigenvalues. Then
\begin{eqnarray}
 \| \rho - \rho' \|_1  = {\rm tr} (| X_{>} \!  - \!  X_{<} |)
 = {\rm tr} ( \sqrt{ X_{>}^2 +  X_{<}^2} )
 = {\rm tr} ( X_{>}  \!  +  \!   X_{<} ) 
 =  {\rm tr} [\phi( X_{>} \!  +  \!  X_{<} ) ], 
\end{eqnarray}
where the last step follows from trace conservation. Because 
$ {\rm tr} (X) = 0$, we have
$ {\rm tr} (  X_{>} ) =  {\rm tr} ( X_< )$
and
$ {\rm tr} [\phi(X_> ) ] =   {\rm tr} [\phi(X_< ) ] .$
Using this leads to
\begin{eqnarray}
 \| \rho - \rho' \|_1  =  2 \,
  {\rm tr} [\phi( X_{>} ) ] .
 \label{result after trace conservation}
\end{eqnarray}

Suppose $A$ and $B$ are positive linear operators, with the eigenvalues of $B$ bounded by unity:
$ 0 \le {\rm spec}(B) \le 1$. In the eigenbasis
$ \{ \KET{\phi_\alpha} \}_{\alpha=1}^d $ of $B$,
such that $ B \KET{\phi_\alpha} = b_\alpha \KET{\phi_\alpha},$ we have
\begin{eqnarray}
B = \sum_\alpha b_\alpha  \KETBRA{\phi_\alpha}, \ \ 0 \le b_\alpha  \le 1.
 \end{eqnarray}
 Evaluating ${\rm tr} (A)$ in the $B$ eigenbasis and using $b_\alpha \le 1 $ leads to
\begin{eqnarray}
{\rm tr} (A) =  \sum_\alpha  \BRA{\phi_\alpha} A
\KET{\phi_\alpha} 
\ge \sum_\alpha b_\alpha \,   \BRA{\phi_\alpha} A
\KET{\phi_\alpha}  = {\rm tr} (AB) .
\label{psd product inequality}
 \end{eqnarray}
 Apply this inequality to (\ref{result after trace conservation}), where for $B$ we choose 
$ B_{\star} \! = \!  {\rm argmax}_{ 0 \preceq B \preceq I}  {\rm tr}(B \phi(\rho - \rho')) $.
Then
\begin{eqnarray}
 \| \rho - \rho' \|_1   = 2 \,
  {\rm tr} [\phi( X_{>} ) ] \ge 2 \, {\rm tr} [ B_{\star}\phi( X_{>} ) ]  \ge 
 2 \, {\rm tr} [ B_{\star}  \phi( X_{>} - X_{<} ) ]  
 =  \|  \phi( \rho - \rho' )  \|_1 ,
\label{result after projection}
\end{eqnarray}
because $ {\rm tr} [ B_{\star}  \phi(X_{<} )  ] $ is nonnegative, proving (\ref{trace distance monotinicity}).


\bibliographystyle{unsrtnat}

\ifUSEBBL
\bibliography{MS2.bbl}
\else
\bibliography{/Users/mgeller/Dropbox/bibliographies/CM,/Users/mgeller/Dropbox/bibliographies/MATH,/Users/mgeller/Dropbox/bibliographies/QFT,/Users/mgeller/Dropbox/bibliographies/QI,/Users/mgeller/Dropbox/bibliographies/group,/Users/mgeller/Dropbox/bibliographies/books}
\fi

\end{document}